\documentclass{article}
\usepackage[preprint]{neurips_data_2024}
\usepackage[most]{tcolorbox}
\usepackage{graphicx}
\usepackage{algorithmic}
\usepackage[linesnumbered,ruled,lined]{algorithm2e}
\usepackage{xcolor,colortbl}
\usepackage{tcolorbox}
\usepackage[nolist,nohyperlinks]{acronym}
\usepackage[utf8]{inputenc}
\usepackage{bm}
\usepackage{lipsum}
\usepackage[inline]{enumitem}
\usepackage{url}
\usepackage{natbib}
\usepackage{multirow}
\usepackage{amsmath}
\usepackage[export]{adjustbox}
\usepackage{balance}
\usepackage{float}
\usepackage{graphicx}
\usepackage[subrefformat=parens,labelformat=parens]{subfig}
\usepackage{svg}
\usepackage{amssymb}
\usepackage{xspace}
\usepackage{wrapfig}
\usepackage{appendix}

\newcommand{\ourdataset}{\textit{mmDoppler}\xspace}

\title{A Dataset for Multi-intensity Continuous Human Activity Recognition through Passive Sensing}

\author{%
  Argha Sen\\ 
  IIT Kharagpur, India\\
  \texttt{arghasen10@gmail.com} \\
  \And
  Anirban Das \\
  NIIT University \\
  \texttt{anirbanfuture@gmail.com} \\
  \And
  Swadhin Pradhan \\
  Cisco Systems, USA \\
  \texttt{swadhinjeet88@gmail.com} \\
  \And
  Sandip Chakraborty \\
  IIT Kharagpur, India \\
  \texttt{sandipc@cse.iitkgp.ac.in} \\
}

\begin{document}

\maketitle

\begin{abstract}
Human activity recognition (HAR) is essential in healthcare, elder care, security, and human-computer interaction. The use of precise sensor data to identify activities passively and continuously makes HAR accessible and ubiquitous. Specifically, millimeter wave (mmWave) radar is promising for passive and continuous HAR due to its ability to penetrate non-metallic materials and provide high-resolution wireless sensing. Although mmWave sensors are effective at capturing macro-scale activities, like exercising, they fail to capture micro-scale activities, such as typing. In this paper, we introduce \ourdataset\footnote{\url{https://github.com/arghasen10/mmdoppler} (Access: \today)}, a novel dataset that utilizes off-the-shelf (COTS) mmWave radar in order to capture both macro and micro-scale human movements using a machine-learning driven signal processing pipeline. The dataset includes seven subjects performing $19$ distinct activities and employs adaptive doppler resolution to enhance activity recognition. By adjusting the radar's doppler resolution based on the activity type, our system captures subtle movements more precisely. \ourdataset includes range-doppler heatmaps, offering detailed motion dynamics, with data collected in a controlled environment with single as well as multiple subjects performing activities simultaneously. The dataset aims to bridge the gap in HAR systems by providing a more comprehensive and detailed resource for improving the robustness and accuracy of mmWave radar activity recognition.

\end{abstract}

\section{Introduction}

Think of the possibilities when technology seamlessly and passively tracks your movements, capturing everything from the subtle flick of a finger to a grand gesture of a hand wave (macro-scale). In recent years, millimeter wave (mmWave) radar technology has made this vision a reality. A mmWave radar's ability to penetrate non-metallic materials, resilience to environmental changes, and high-resolution sensing capabilities distinguish it from other radar technologies. Traditional methods of human activity recognition (HAR) rely on sensors like LiDAR or depth cameras, which excel at capturing large movements such as walking or running~\cite{alam2020lamar, alam2021palmar, palipana2021pantomime, singh2019radhar}. However, these methods often fail to detect more nuanced activities like typing or subtle hand movements. Therefore, advanced sensor techniques are necessary for accurately capturing both macro- and micro-scale human movements.

To address this challenge, we introduce \ourdataset, an innovative dataset of human activities collected using a Commercial off-the-shelf (COTS) mmWave radar. This dataset bridges the gap between macro and micro-scale activity recognition by capturing activities through their \textit{range-doppler} signatures, along with \textit{point cloud} data. A point cloud represents a collection of three-dimensional points reflecting radar signals from objects in the environment, providing detailed spatial information. The range-doppler heatmaps are 2D image representations where the abscissa represents the range (location of human subjects), and the ordinate indicates the doppler speed at which subjects are moving. These heatmaps combine range and doppler information from radar signals, offering a detailed view of motion dynamics over time. By integrating these data types, \ourdataset encompasses a wide range of activities—from large actions like walking and squats to fine-grained behaviors like typing and sitting. Seven subjects performed 19 different activities in 3 unique indoor settings, providing a rich and diverse dataset for advancing HAR research.


Capturing the subtle movements of micro-scale activities even with mmWave radar presents a significant challenge. To overcome this, we utilize an innovative approach with adaptive doppler resolution instead of capturing only the pointcloud dataset as done in the prior works~\cite{alam2020lamar, palipana2021pantomime, singh2019radhar}. Higher doppler resolution allows for precise detection of small movement changes, while lower doppler resolution is sufficient for detecting larger movements. However, higher doppler resolution also introduces more noise. Therefore, by optimizing doppler resolution based on the activity type, our system effectively captures the nuances of various movements for \ourdataset.

For macro-scale activities characterized by significant movement, such as walking or running, a lower doppler resolution is used to efficiently capture the prominent velocity changes without needing fine detail. Conversely, for micro-scale activities involving minimal movement, like typing or subtle hand gestures, a higher doppler resolution is employed, allowing the radar to detect subtle velocity changes with greater detail and accuracy. This adaptive doppler resolution approach facilitates the robustness and precision of HAR systems, making them more effective across a wide range of applications that require both macro and micro-scale movement recognition.

That said, \ourdataset can be characterized by the following key features:

    \noindent
$\bullet\;$ \textbf{Range-doppler information:} \ourdataset is specifically curated with range-doppler heatmaps of different activities of daily leaving. This is in contrast to the traditional activity signature datasets with only point clouds generated by mmWave radar. 

    \noindent
$\bullet\;$ \textbf{Adaptive Doppler resolution:} as we observe that the macro and micro-scale activities can be better captured by dynamically switching the radar sensitivity. \ourdataset, therefore, is curated with adaptive doppler resolution.

    \noindent
$\bullet\;$ \textbf{Diverse Activity Set:} We have included a wide array of 19 different activity classes from Activities of Daily Living (ADLs), Instrumental Activities of Daily Living (IADLs)~\cite{adlsiadls}, and daily indoor exercises – (i) macro activities like walking, running, jumping, clapping, lunges, squats, waving, vacuum cleaning, folding clothes, changing clothes, and (ii) micro activities like laptop-typing, phone-talking, phone-typing, sitting, playing guitar, eating food, combing hair, brushing teeth, and drinking water to ensure comprehensive coverage of daily human behaviors. 

    \noindent
$\bullet\;$ \textbf{Real-world setups:} The data is collected mimicking diverse scenarios of real indoor environments with the presence of furniture and other objects usually found in an indoor setup. This promotes the real-world applicability of any system evaluated on this dataset. 

    \noindent
$\bullet\;$\textbf{Controlled Environment:} To avoid unaccountable noise signatures, Data collection was conducted in a controlled indoor environment. This also ensures the quality and consistency of the data. \ourdataset comprises an effective duration of $23100$ seconds of data, $27\%$ more than the closest largest dataset in existence~\cite{cui2024milipoint}.

    \noindent
$\bullet\;$ \textbf{Multiple Subjects:} \ourdataset uniquely includes multiple subjects performing various activities simultaneously, unlike traditional datasets that typically focus on single-subject activities~\cite{singh2019radhar, palipana2021pantomime, ahuja2021vid2doppler}. This feature allows our dataset to account for the complexity and variability of real-world scenarios where interactions between multiple individuals are common.


As a result, our dataset provides a valuable resource for researchers and developers working on HAR systems for continuous passive sensing. Combining adaptive doppler resolution and range-doppler heatmaps enables more accurate and robust activity recognition across a wide range of activities. By combining mmWave radar technology with point cloud data, this dataset overcomes the limitations of previous HAR systems solely based on point clouds. Making this dataset public will enable further research and development of innovative solutions that leverage the capabilities of mmWave radar to comprehensively and reliably detect continuous human activity through passive sensing.
\section{Related Work}
Human Activity (HA) detection has been extensively studied over the past few decades. However, the methods and technologies used for assessment have evolved significantly in response to technological advancements. Camera-based~\cite{kumrai2020human} approaches have been leveraged for quite some time. However, despite being effective, they primarily suffer from privacy concerns or poor lighting conditions. Studies have used wearables for active sensing to detect human activity~\cite{lawal2019deep, sandhu2021solar}. However, these methods are not necessarily seamless or pervasive. Alternative passive sensing methods, such as those based on acoustics and radio frequency, offer promising solutions. Passive acoustic sensing~\cite{wang2018c, li2022lasense, wang2023df} uses audio chirps detected by microphones to study activities but is susceptible to environmental noise and interference~\cite{jiang2018towards}. Radio-frequency (RF) sensing, using WiFi~\cite{ding2020rf}, RFIDs~\cite{kellogg2014bringing}, and UWB radars~\cite{zhang2022mobi2sense}, captures human dynamics by analyzing changes in radio waves~\cite{chen2018wifi}. WiFi Channel State Information (CSI) has been widely explored~\cite{cominelli2023exposing, soltanaghaei2020robust, tan2019multitrack} but is complex and affected by environmental factors~\cite{chen2023cross, chen2023wider}. FMCW techniques with specialized hardware offer higher resolution but are costly~\cite{fan2020home, li2019making}. COTS mmWave sensors, like IWR1642~\cite{iwr1642boost}, provide better range resolution and can detect fine movements~\cite{jiang2018towards, rao2017introduction}. Studies using mmWave sensors rely on emitted chirps to capture activity signatures~\cite{xie2023mm3dface, singh2019radhar, lu2020see, shuai2021millieye, wang2021m, bhalla2021imu2doppler}, with features like point clouds and range-doppler proving effective for movement assessment~\cite{ahuja2021vid2doppler, sen2023mmassist, sen2023mmdrive}. While WiFi and UWB-based methods face various challenges, mmWave sensors offer a robust and potentially fine-grained solution for activity assessment.

Over the past few years, numerous datasets have been developed to support human activity detection research, utilizing various sensing modalities. Traditional datasets often rely on wearable sensors~\cite{NEURIPS2022_5985e81d} and video data~\cite{zheng2024ha} to capture activities. Dataset using a depth camera~\cite{chao2022czu} instead of an RGB camera, can also be seen in the existing works. Similarly, IMU sensors (such as accelerometers, gyroscopes, etc.) using inertial measurement units to detect various physical activities have been used to curate dataset~\cite{bockwear, wc78-mt44-23} for human activity assessment.  Acoustic sensing datasets~\cite{JUNG2020103177}, on the other hand, employ microphones to detect activities through sound. RFID-based datasets~\cite{a3e9-f387-23} capture activities by tracking RFID tag movements. WiFi Channel State Information (CSI)~\cite{wc78-mt44-23}, sometimes with multimodal information~\cite{yang2024mm}, has also been used to capture the impact of human movements on WiFi signals. Despite their utility, these datasets often have limitations such as privacy concerns, poor lighting conditions, acoustic noise, and the need for participants to wear or interact with devices constantly.

More recently, datasets generated by mmWave radar, more specifically, FMCW mmWave radars, have gained attention due to their non-intrusive nature and high resolution. These datasets capture fine-grained details of human movements by analyzing the frequency shifts and distances of reflected radar signals. As mentioned earlier, these datasets can be categorized primarily into two types: point clouds and range-doppler heat maps. Within the first category, multiple point cloud~\cite{yang2024mm, cui2024milipoint, singh2019radhar, an2022mri} datasets exist for studying both macro and micro-level human activities or gestures. Interestingly, these datasets primarily focus on single-subject activities. However, the datasets for human activity assessment that capture range-doppler information are not available to the best of our knowledge. We further compare the available mmWave-based datasets with \ourdataset in terms of relevant features in Table~\ref{tab:comparison}. Interestingly, recent studies~\cite{sen2023continuous} have shown that range-doppler heatmaps are significantly effective over point clouds for activity assessment. As a result of using range-doppler information for human activity detection, new avenues are opened for unobtrusive and continuous monitoring, enhancing both the scope and accuracy of the recognition of human activity.

\begin{table}
	\centering
	\scriptsize
	\caption{Comparisons of \ourdataset with published datasets.}
	\label{tab:comparison}
	\begin{tabular}{|l|l|l|l|l|l|l|l|} 
		\hline
		Datasets           & Modality                                                                                                        & Activity type                                                                         & \# Classes  & Granularity                                                                       & \# Frames    & \begin{tabular}[c]{@{}l@{}}Effective \\Duration (s)\end{tabular} & \begin{tabular}[c]{@{}l@{}}Multi\\Subjects\end{tabular}  \\ 
		\hline
		mRI~               & \begin{tabular}[c]{@{}l@{}}mmWave pointcloud, \\ RGB, Depth Camera,\\ and IMU signals.\end{tabular}             & Pose estimation                                                                       & 12          & Macro scale                                                                       & 160k         & 5333                                                             & No                                                       \\ 
		\hline
		mm-Fi~             & \begin{tabular}[c]{@{}l@{}}RGB, Depth camera, \\ LiDAR pointcloud,\\ mmWave pointcloud,\\ WiFi CSI\end{tabular} & Daily activities                                                                      & 27          & Macro scale                                                                       & 320k         & 10666                                                            & No                                                       \\ 
		\hline
		milipoint~         & mmWave pointcloud                                                                                               & Daily activities                                                                      & 49          & Macro scale                                                                       & 545k         & 18166                                                            & No                                                       \\ 
		\hline
		RadHAR~            & mmWave pointcloud                                                                                               & Exercise                                                                              & 5           & Macro scale                                                                       & 167k         & 5566                                                             & No                                                       \\ 
		\hline
		\textbf{\ourdataset} & \begin{tabular}[c]{@{}l@{}}\textbf{mmWave pointcloud,}\\\textbf{Range-Doppler}\\\textbf{heatmaps}\end{tabular}  & \begin{tabular}[c]{@{}l@{}}\textbf{Daily activities,}\\\textbf{Exercise}\end{tabular} & \textbf{19} & \begin{tabular}[c]{@{}l@{}}\textbf{Macro~and }\\\textbf{micro scale}\end{tabular} & \textbf{75k} & \textbf{23100}                                                   & \textbf{Yes}                                             \\
		\hline
	\end{tabular}
\end{table}
\section{Dataset}

\begin{figure}[htbp]
    \centering
    \subfloat[Map representation of R1]{\includegraphics[width=0.3\textwidth]{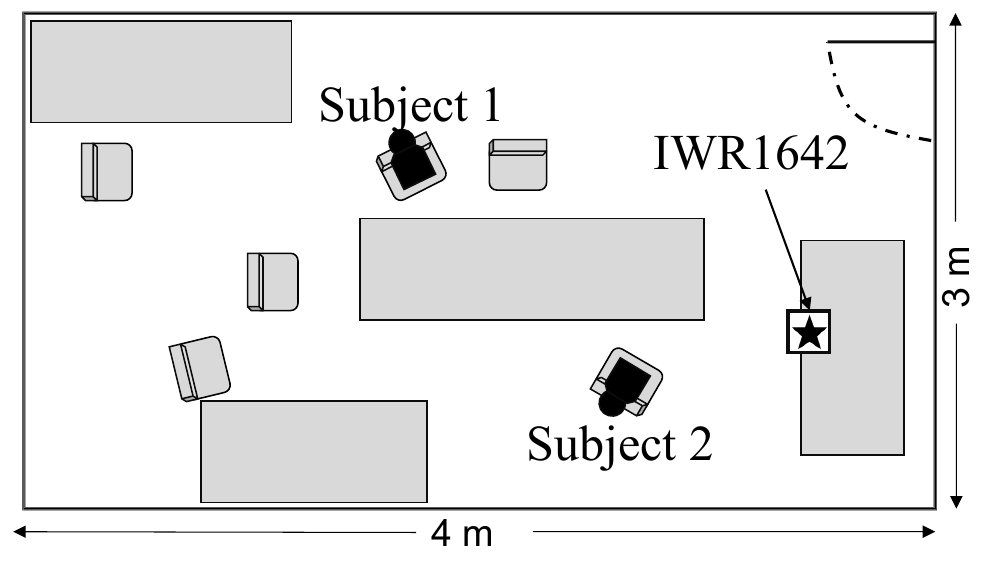}\label{fig:room11}}
    \hfill
    \subfloat[Map representation of R2]{\includegraphics[width=0.3\textwidth]{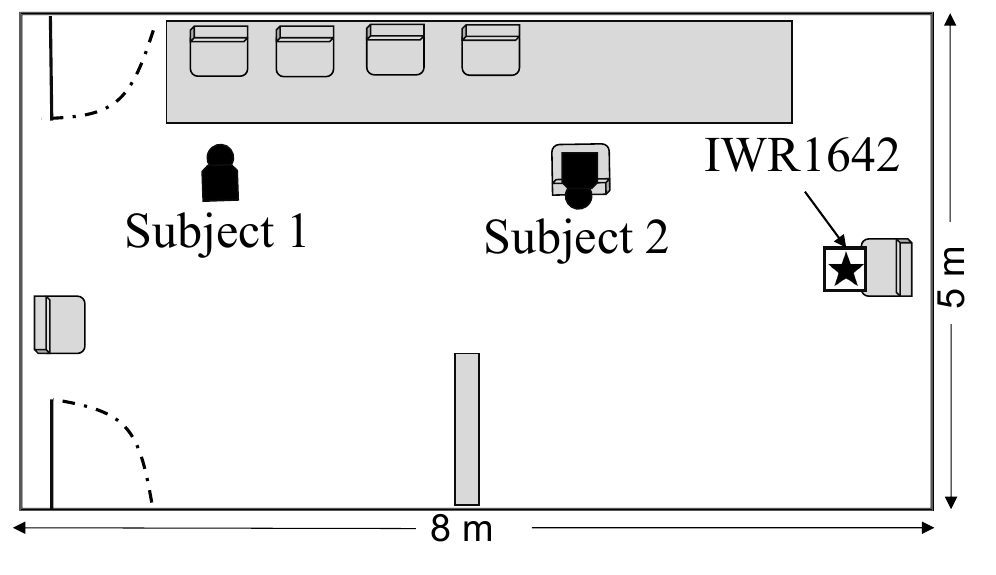}\label{fig:room21}}
    \hfill
    \subfloat[Map representation of R3]{\includegraphics[width=0.34\textwidth]{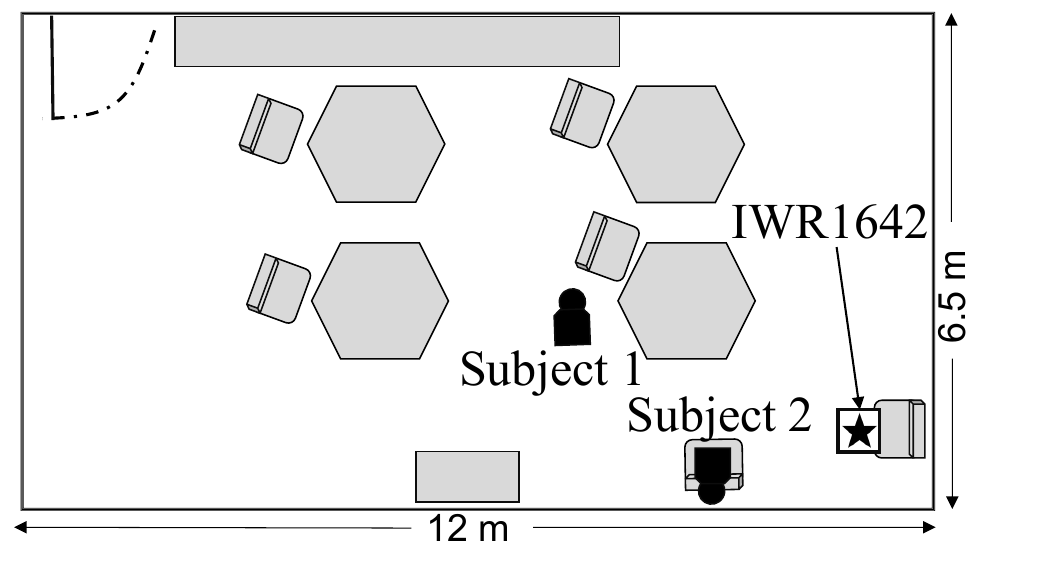}\label{fig:room31}} \\
    
    \subfloat[Activity capture scenario in R1]{\includegraphics[width=0.3\textwidth]{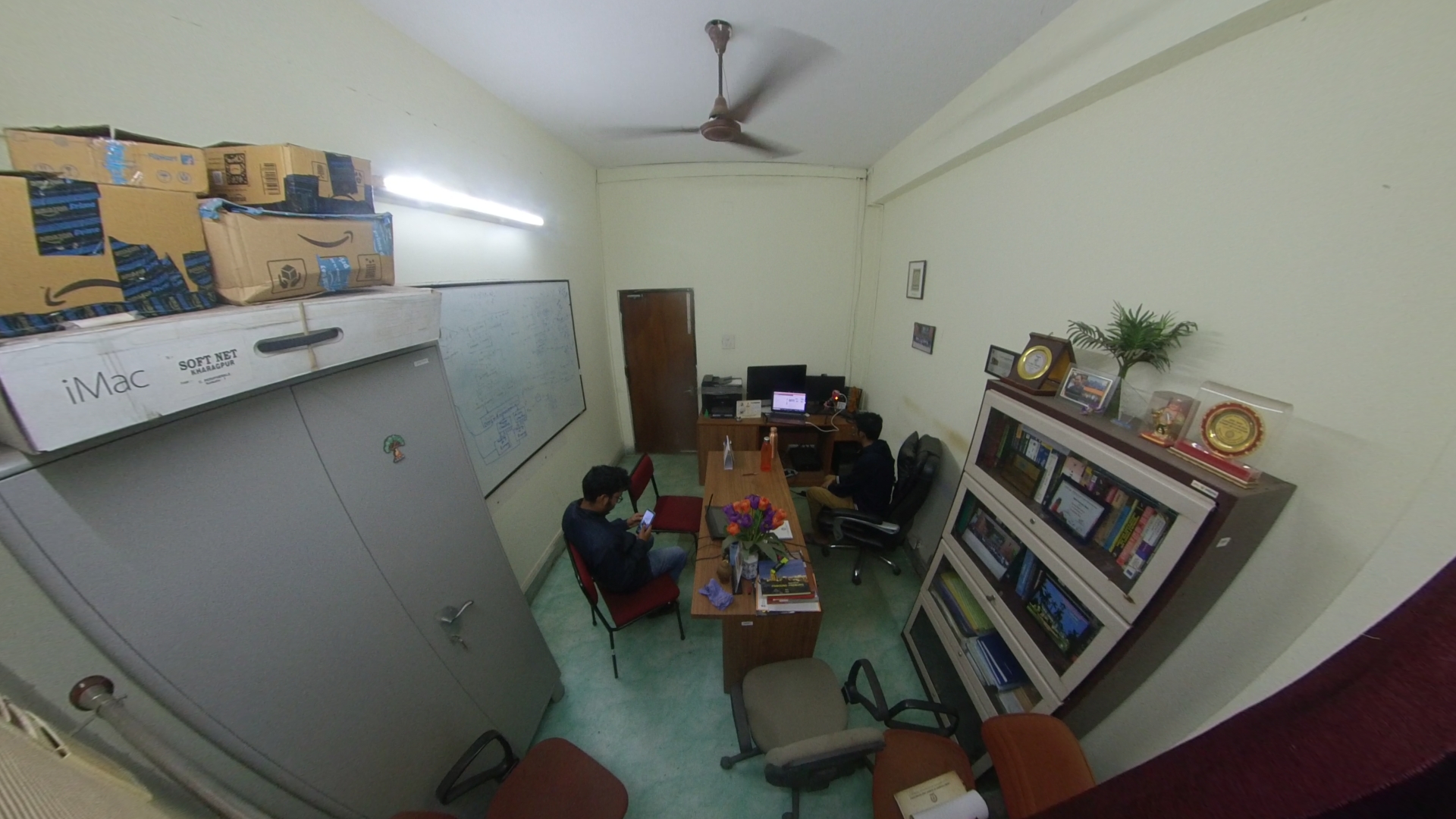}\label{fig:r1setup}}
    \hfill
    \subfloat[Activity capture scenario in R2]{\includegraphics[width=0.3\textwidth]{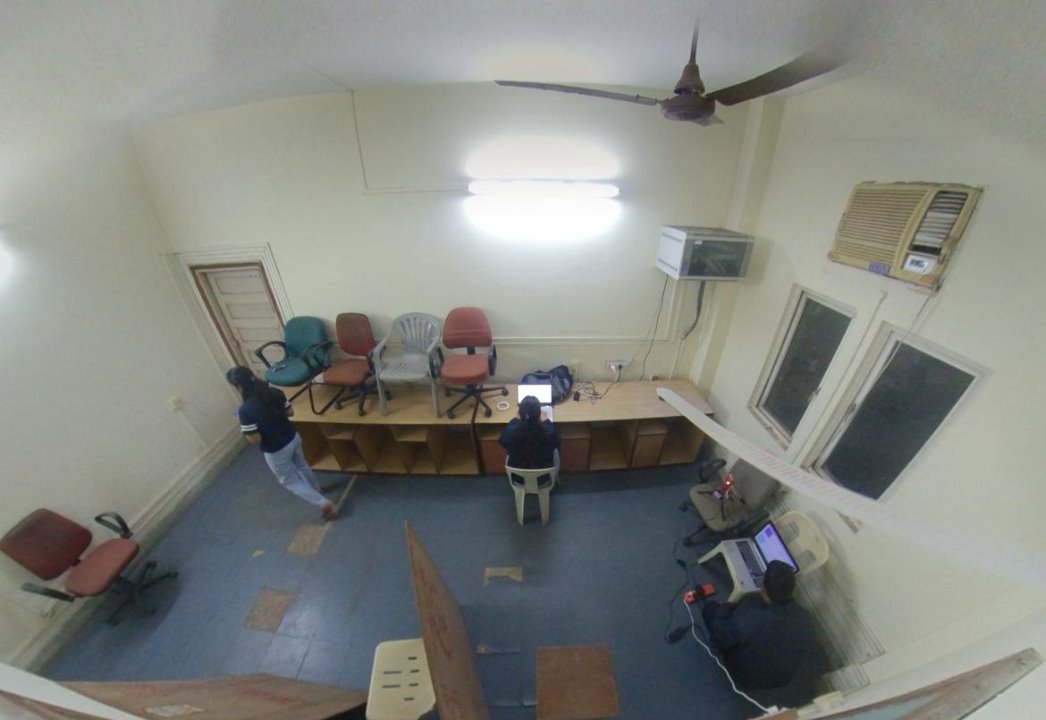}\label{fig:r2setup}}
    \hfill
    \subfloat[Activity capture scenario in R3]{\includegraphics[width=0.3\textwidth]{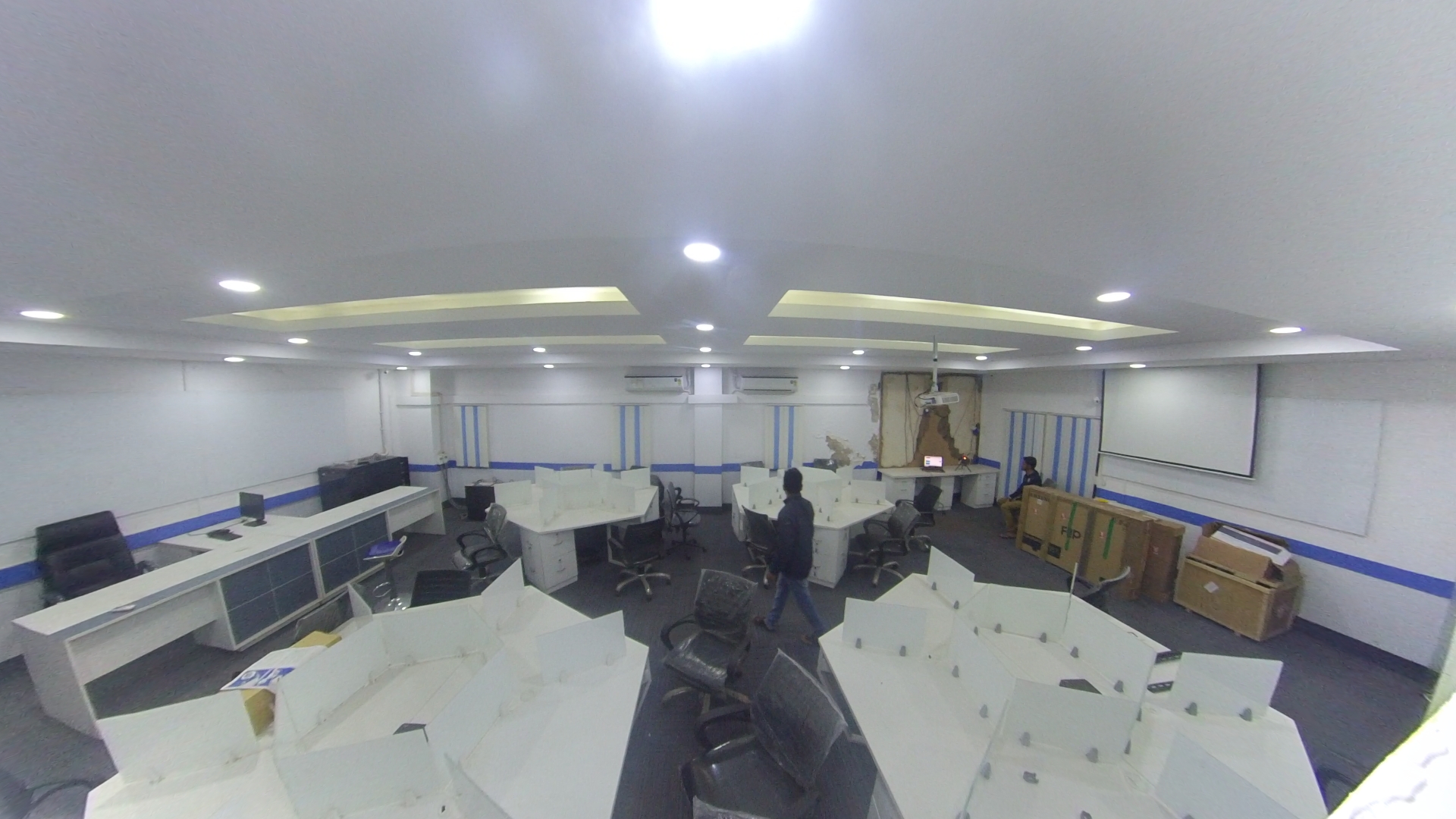}\label{fig:r3setup}} \\
    
    \subfloat[The setup for data collection]{\includegraphics[width=0.3\textwidth]{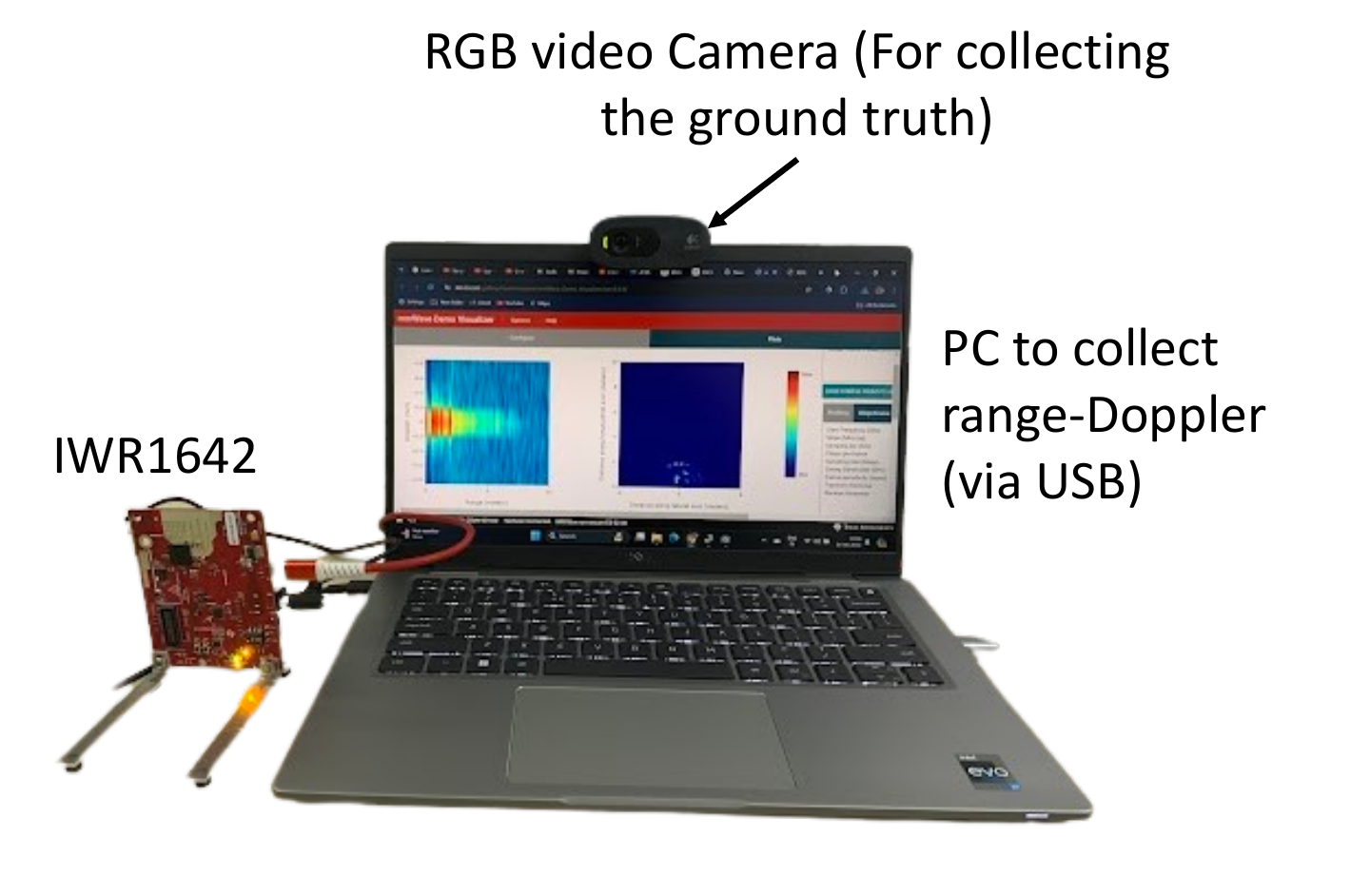}\label{fig:datacollsetup}}
    \hspace{30pt}
    \subfloat[A visual representation of simultaneous multiple activity capturing]{\includegraphics[width=0.3\textwidth]{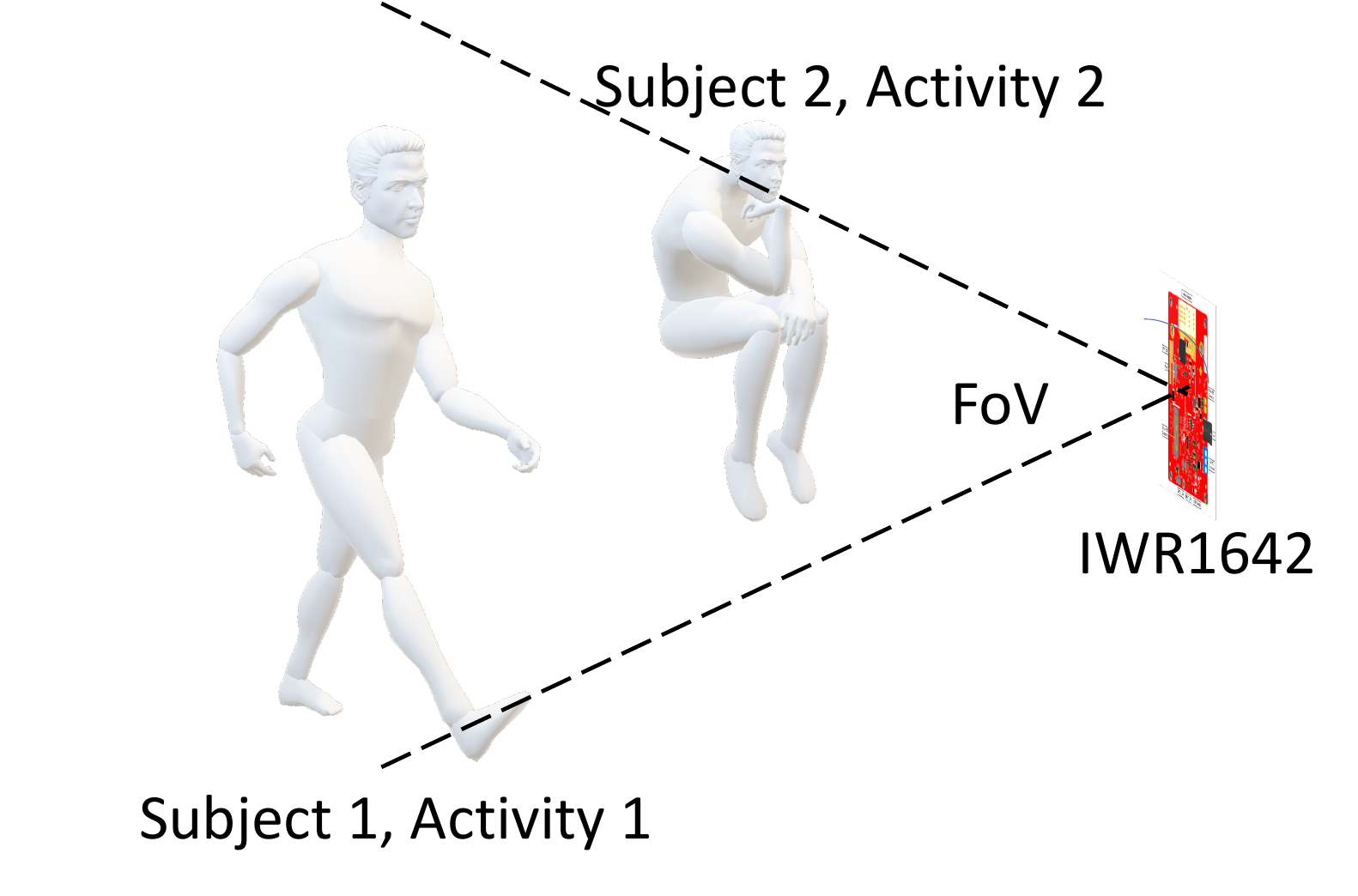}\label{fig:datacoll}}
    
    \caption{IWR1642 hardware based setup and data collection in different rooms (R1, R2, R3).}
    \label{fig:implementation_setup}
\end{figure}

As shown in \figurename~\ref{fig:implementation_setup}, our data collection setup is built on a Commercial off-the-shelf (COTS) millimeter wave radar, the \texttt{IWR1642BOOST}~\cite{iwr1642boost}. The system was tested in three distinct rooms to capture a variety of environments:
\begin{itemize}
    \item \textbf{R1:} An office cabin of size $4 \times 3$ m$^2$
    \item \textbf{R2:} A classroom of size $8 \times 5$ m$^2$
    \item \textbf{R3:} A laboratory of size $12 \times 6.5$ m$^2$
\end{itemize}
Refer to \figurename~\ref{fig:implementation_setup}(d), \ref{fig:implementation_setup}(e), and \ref{fig:implementation_setup}(f) for the specific room setups.

The ground truth for each subject's activity was meticulously annotated with the help of video footage captured using a USB camera. This ensured accurate labelling and high-quality data. The sensor data are transmitted via a USB cable at a baud rate of 921600 to a Dell Inspiron Laptop having Intel(R) Core(TM) i7-8550U CPU with 16 GB of RAM. 

\begin{figure*}
    \centering
    \subfloat[]
    {\includegraphics[width=0.45\textwidth]{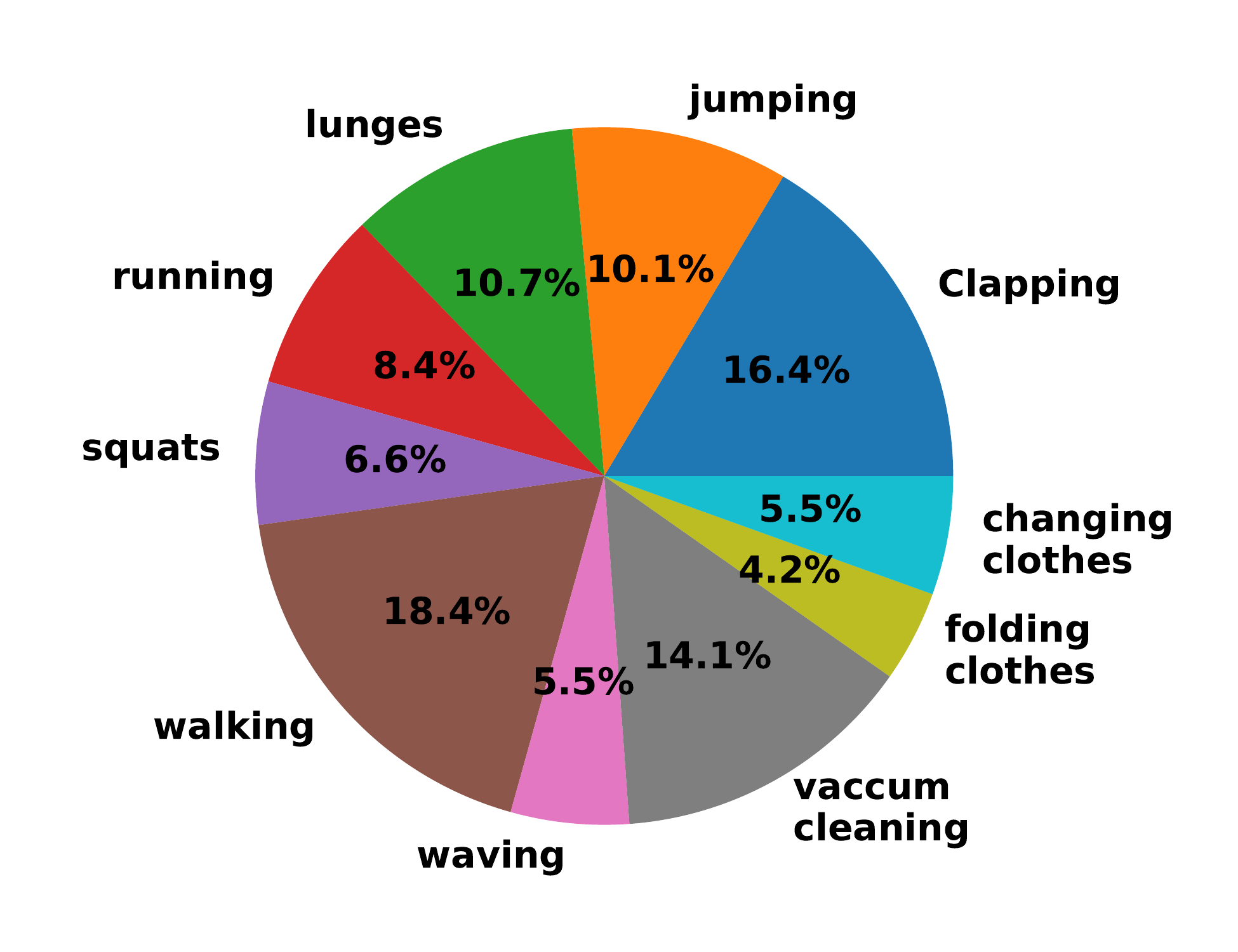}\label{fig:macro_dist}}
     \subfloat[]{\includegraphics[width=0.45\textwidth]{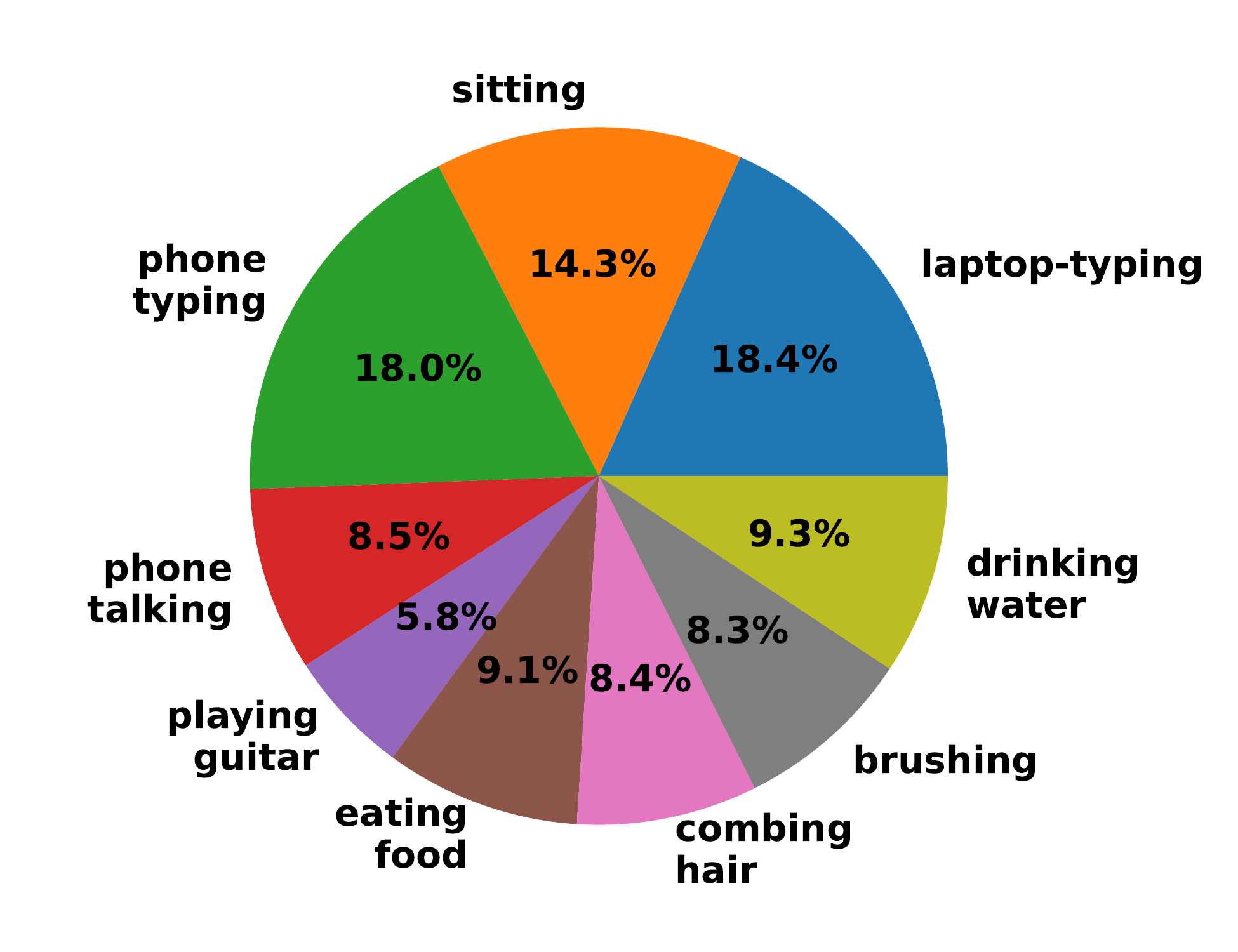}}\label{fig:micro_dist}
     \caption{Dataset Distribution per activities (in $\%age$) (a) Macro Activities, (b) Micro Activities.}
\end{figure*}

\subsection{Radar Configuration}
The IWR1642BOOST radar is configured with two transmitter and four receiver antennas, operating at frequencies ranging from $77$ to $81$ GHz, providing a bandwidth of $4$ GHz. We have employed two distinct radar configurations tailored for different use cases: macro-scale activities and micro-scale activities, as detailed in \tablename~\ref{tab:radar_conf}.

For macro-scale activities, the radar configuration is optimized to capture larger movements. We set the frame periodicity to $200$ milliseconds, resulting in a frame rate of $5$ frames per second (FPS). This configuration enables the transmission of larger range-Doppler heatmaps, with a matrix size of $16 \times 256$, via USB. The range resolution achieved is $4.36$ cm, with a maximum unambiguous range of $9.02$ meters. Additionally, this setup can measure a maximum radial velocity of $1$ m/s, with a doppler resolution of $0.13$ m/s. The sensor transmits $32$ chirps per frame, which is suitable for capturing the dynamics of macro-scale activities.

For micro-scale activities, where subtle movements need to be accurately detected, we employ a higher Doppler resolution. The Doppler resolution is set to $0.01$ m/s to ensure precise detection of minimal movements. The range-doppler heatmap for this configuration is of size $128 \times 64$, with $128$ number of chirps, supporting a frame rate of $2$ FPS. This setup is particularly effective for capturing fine-grained details of micro-scale activities, such as typing or subtle micro-movements.

These configurations, with their respective range and Doppler resolutions, are meticulously designed to optimize the radar's performance for different scales of activity, ensuring accurate and robust data collection for our HAR dataset.

\begin{table}
	\centering
	\scriptsize
	\caption{Radar configuration}
	\label{tab:radar_conf}
	\begin{tabular}{|c|c|c|} 
		\hline
		\textbf{Parameters}                    & \textbf{Macro} & \textbf{Micro}  \\ 
		\hline
		\textbf{Start Frequency}               & \multicolumn{2}{c|}{77 GHz}      \\ 
		\hline
		\textbf{End Frequency}                 & \multicolumn{2}{c|}{81 GHz}      \\ 
		\hline
		\textbf{Range Resolution (cm)}         & 4.36           & 12.5            \\ 
		\hline
		\textbf{Maximum Range(m)}              & 9.02           & 6.4             \\ 
		\hline
		\textbf{Maximum Radial Velocity (m/s)} & 1              & 0.64            \\ 
		\hline
		\textbf{Velocity Resolution (m/s)}     & 0.13           & 0.01            \\ 
		\hline
		\textbf{Azimuthal Resolution (Degree)} & \multicolumn{2}{c|}{14.5°}       \\ 
		\hline
		\textbf{Frames per Second}             & 5              & 2               \\ 
		\hline
		\textbf{Chirps Per Frame}              & 32             & 64              \\ 
		\hline
		\textbf{ADC Samples per Chirp}         & \multicolumn{2}{c|}{256}         \\
		\hline
	\end{tabular}
\end{table}

\subsection{Data Collection Setup}\label{sec:data_collection}
Data collection was conducted with $7$ subjects ($3$ female and $4$ male), aged between $23$ and $35$. The total duration of data collection spanned $7$ hours, covering $19$ different activity classes, encompassing both macro and micro activities. Specifically, we collected $49$k macro range-doppler samples and $25$k micro range-doppler samples, along with $75$k pointcloud data samples.

The majority of the data was collected in a controlled environment where subjects were instructed to perform specific activities. Additionally, to introduce variability and capture real-world scenarios, we included some sessions where subjects could choose and perform any activity from the predefined set in an uncontrolled fashion. This dual approach allowed us to experiment with both controlled and in-the-wild setups, enhancing the dataset's robustness and applicability.

We utilized the \textit{mmWave-Demo-Visualizer} tool from Texas Instruments~\cite{tiMmWaveDemo}, with a custom patch implemented to extract raw data, including range-Doppler heatmaps, under different radar configurations. Human activities were meticulously annotated using video footage captured by an additional USB camera installed in the room. Two volunteers assisted in annotating the activities, ensuring high accuracy and reliability of the labels. As a result of this comprehensive data collection methodology, a diverse and well-annotated dataset is provided for development and evaluation of HAR systems using a COTS mmWave radar.

\subsection{License and Consent}
The dataset is free to download and can be used with GNU General Public License for non-commercial purposes. The participants received \$20 per session as remuneration during the data collection. The experimental procedures were approved by the Ethical Review Committee at IIT Kharagpur, India, with the Approval Number: IIT/SRIC/DEAN/2023, dated July
31, 2023, under the study title: “Human Activity Monitoring in Pervasive Sensing Setup".

\section{Observations and Benchmark}\label{sec:benchmark}
We consider $19$ different activity classes from \textit{Activities of Daily Living} (ADLs), \textit{Instrumental Activities of Daily Living} (IADLs)~\cite{adlsiadls}, and \textit{daily indoor exercises} -- (i) \textit{macro activities} like walking, running, jumping, clapping, lunges, squats, waving, vacuum cleaning, folding clothes, changing clothes, and (ii) \textit{micro activities} like laptop-typing, phone-talking, phone-typing, sitting, playing guitar, eating food, combing hair, brushing teeth, and drinking water. In contrast to the existing literature that primarily uses voxelized pointcloud~\cite{singh2019radhar, wang2021m, cai2023millipcd} or 1D doppler~\cite{ahuja2021vid2doppler, bhalla2021imu2doppler}, in this paper, we explore range-doppler 2D heatmaps for activity classification; the primary motive is to find a parameter that can detect both macro and micro activities simultaneously from different users. 

\begin{figure*}[t]
    \def\wide{0.19}
    \subfloat[Clapping]{
    \includegraphics[trim={22mm 0 0 7mm}, width=\wide\columnwidth]{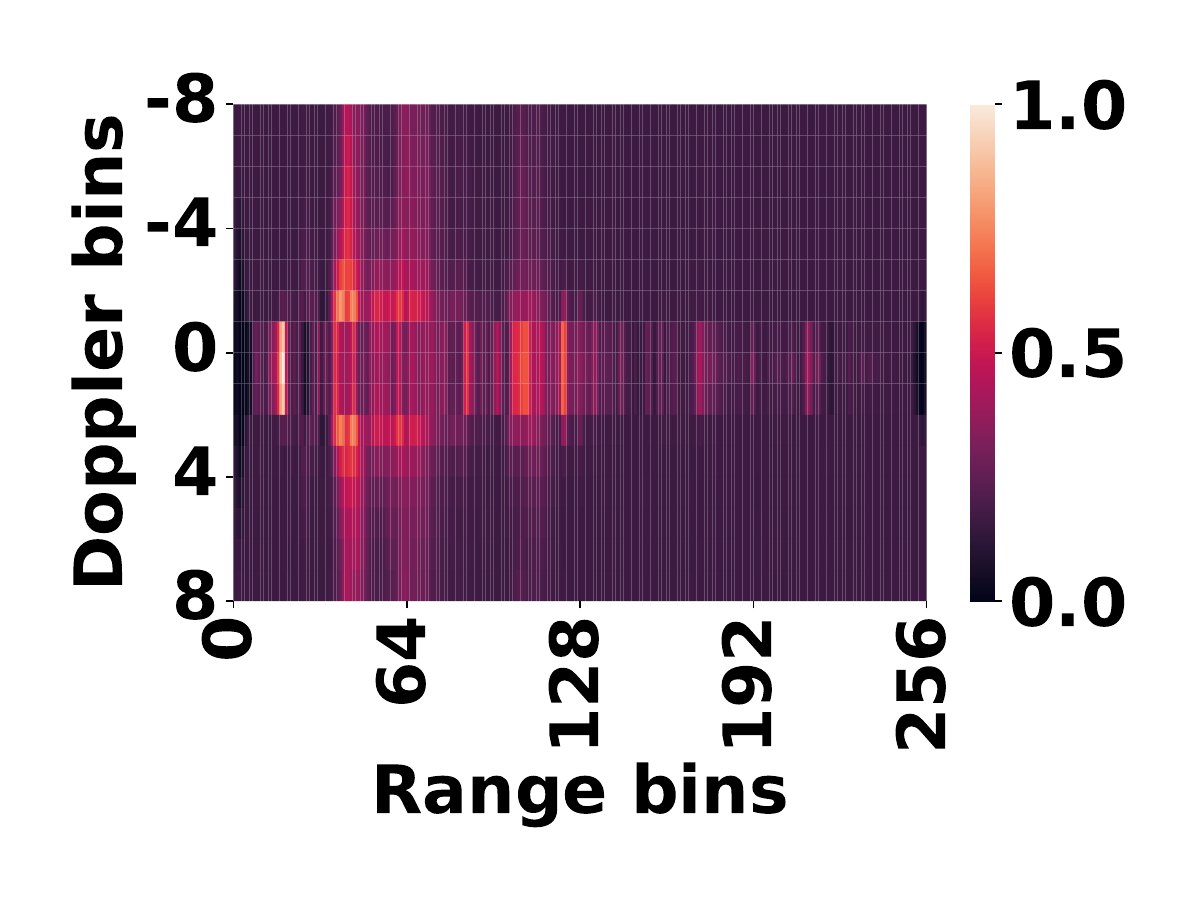}
    }
        \subfloat[Jumping]{
    \includegraphics[trim={22mm 0 0 7mm},width=\wide\columnwidth]{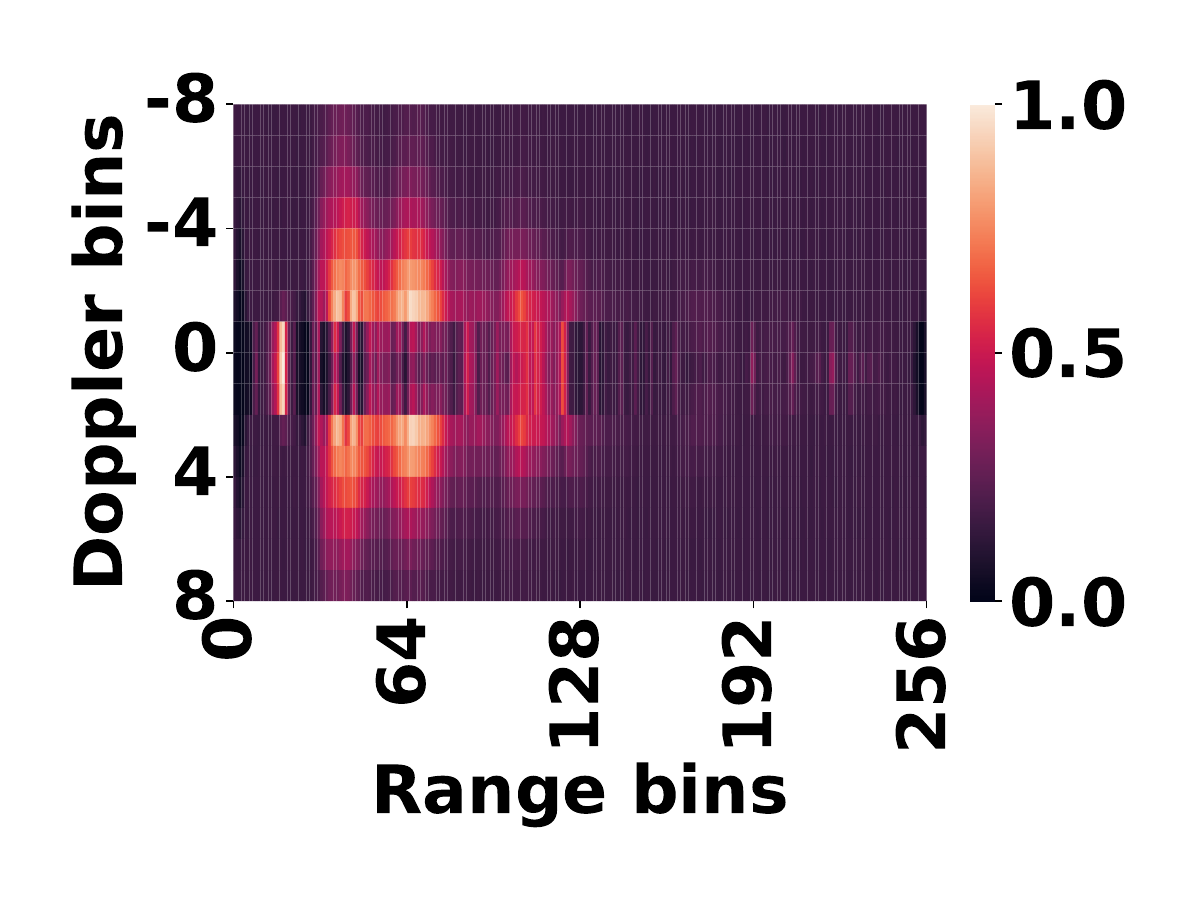}
    }
        \subfloat[Lunges]{
    \includegraphics[trim={22mm 0 0 7mm},width=\wide\columnwidth]{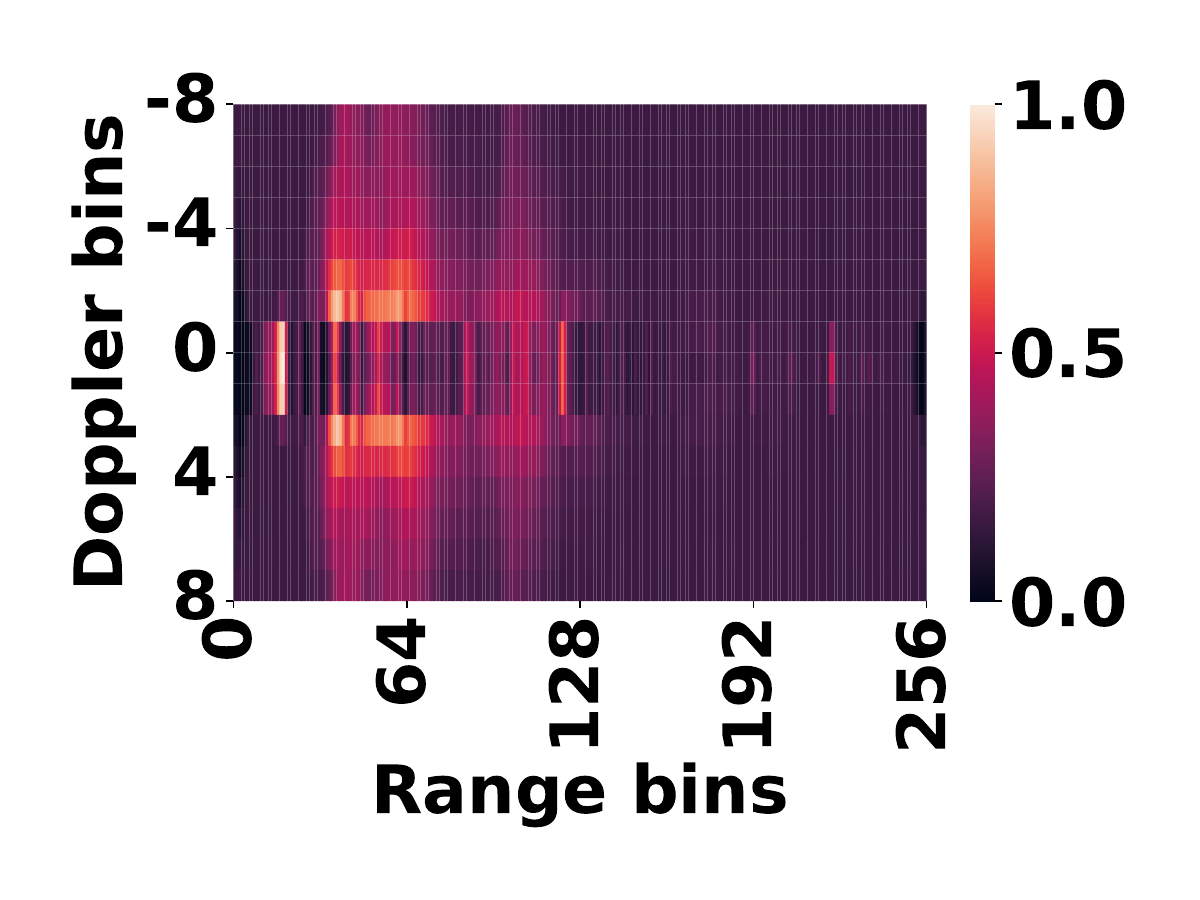}
    }
        \subfloat[Walking]{
    \includegraphics[trim={22mm 0 0 7mm},width=\wide\columnwidth]{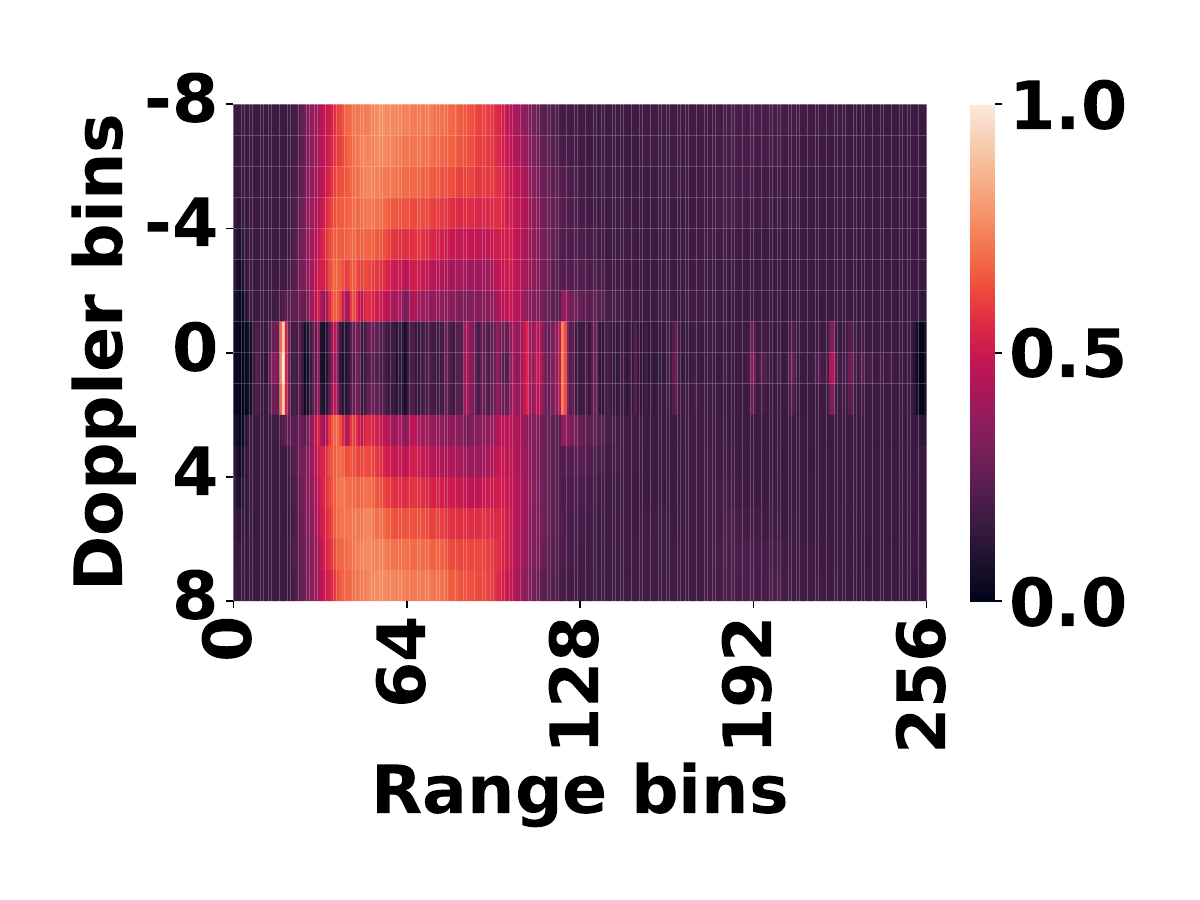}
    }
        \subfloat[Squats]{
    \includegraphics[trim={22mm 0 0 7mm},width=\wide\columnwidth]{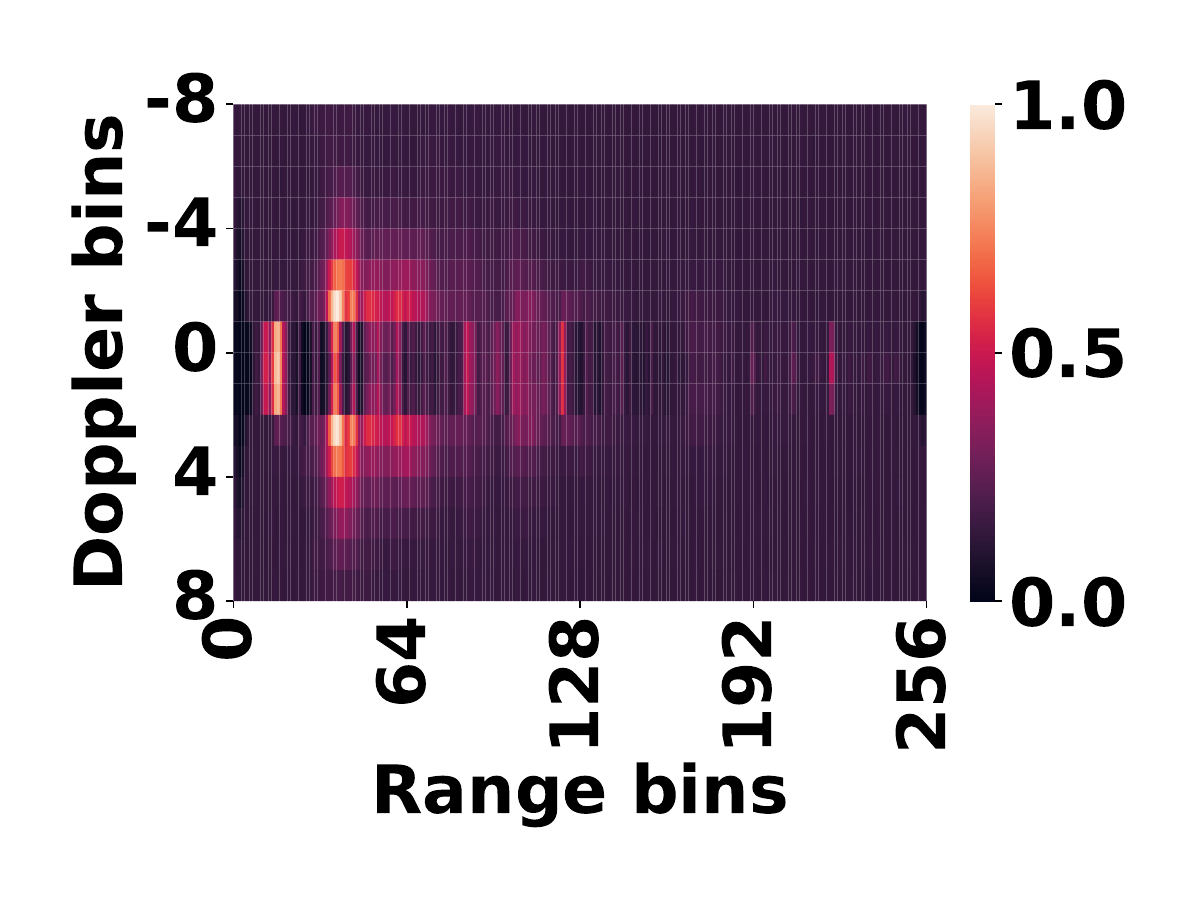}
    }\\
    \vspace{-5pt}
        \subfloat[Waving]{
    \includegraphics[trim={22mm 0 0 7mm},width=\wide\columnwidth]{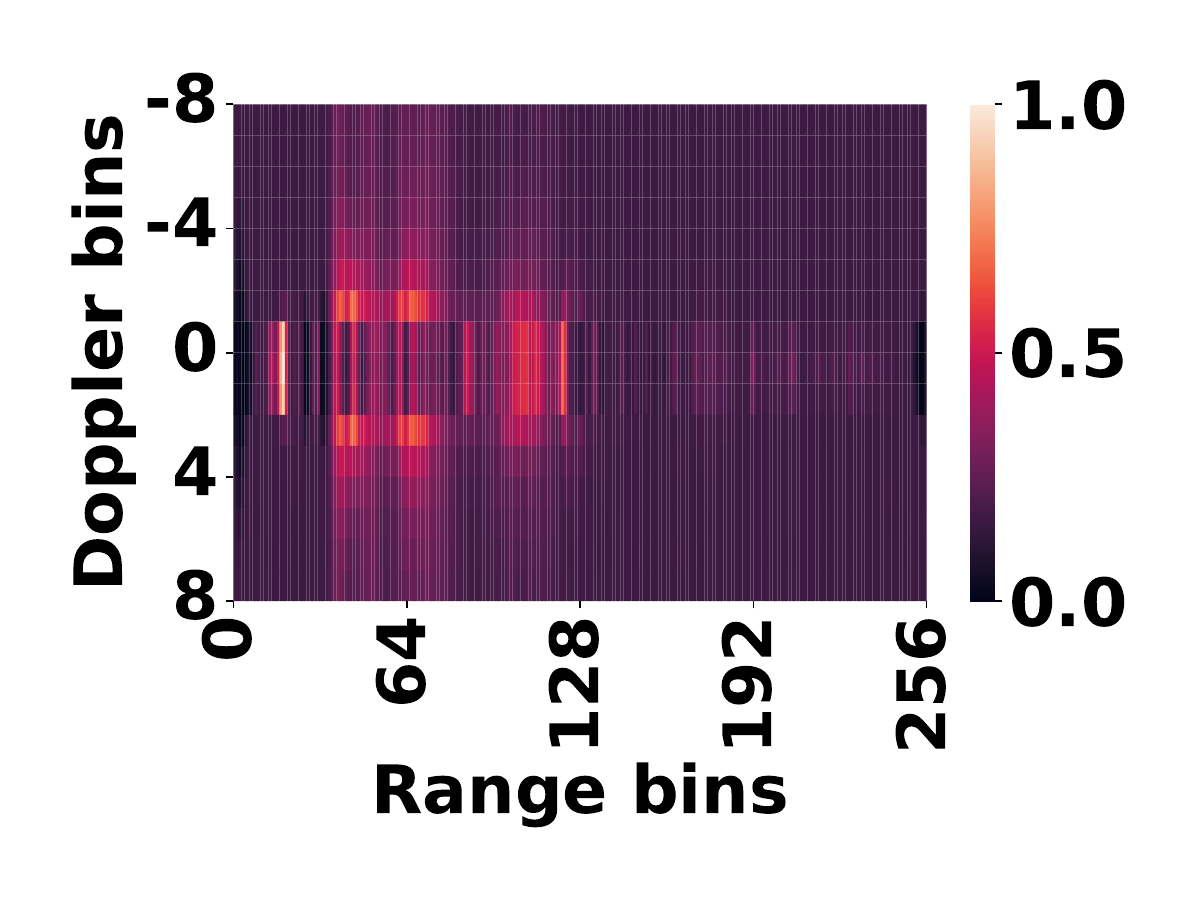}
    }
        \subfloat[Fold \\Cloth]{
    \includegraphics[trim = {22mm 0 0 0mm}, width=\wide\columnwidth]{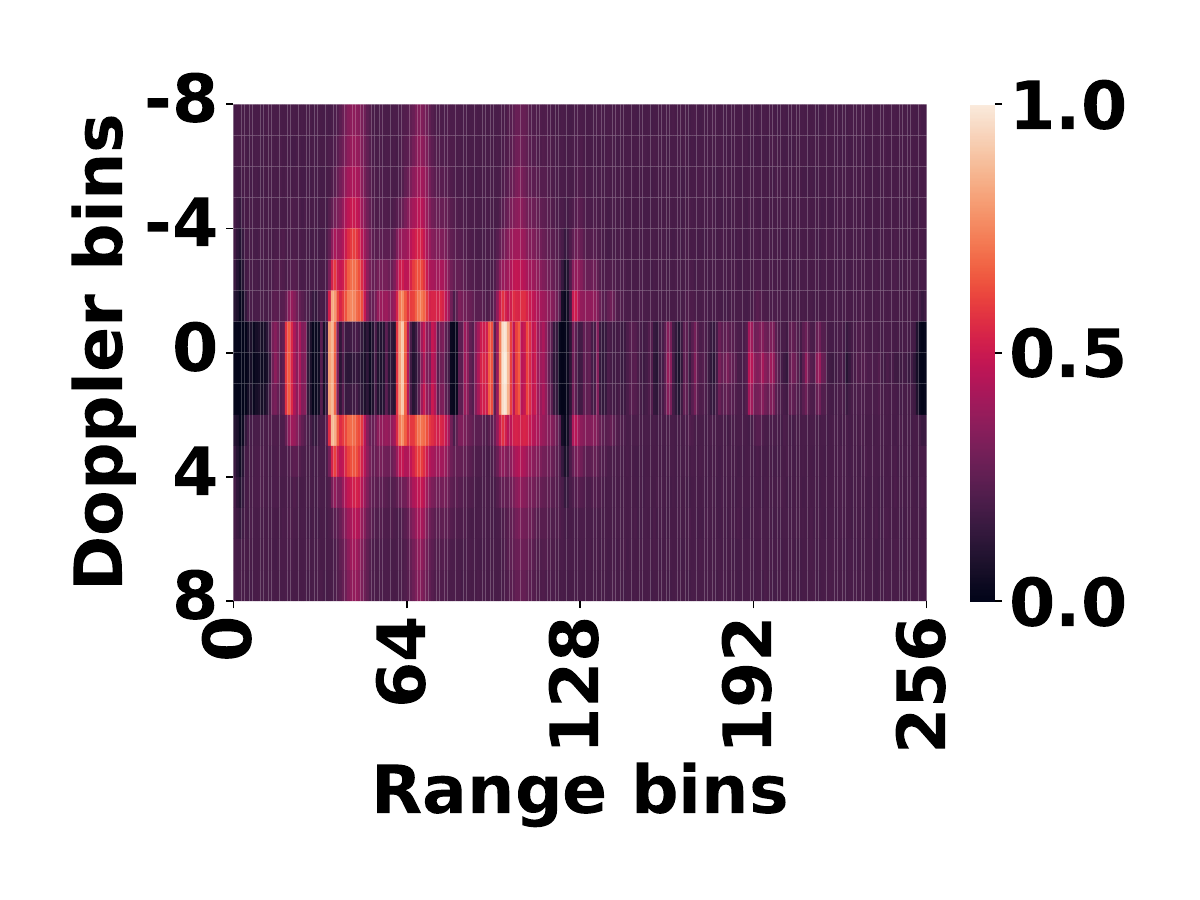}
    }
    \subfloat[Change \\Cloth]{
    \includegraphics[trim = {22mm 0 0 7mm}, width=\wide\columnwidth]{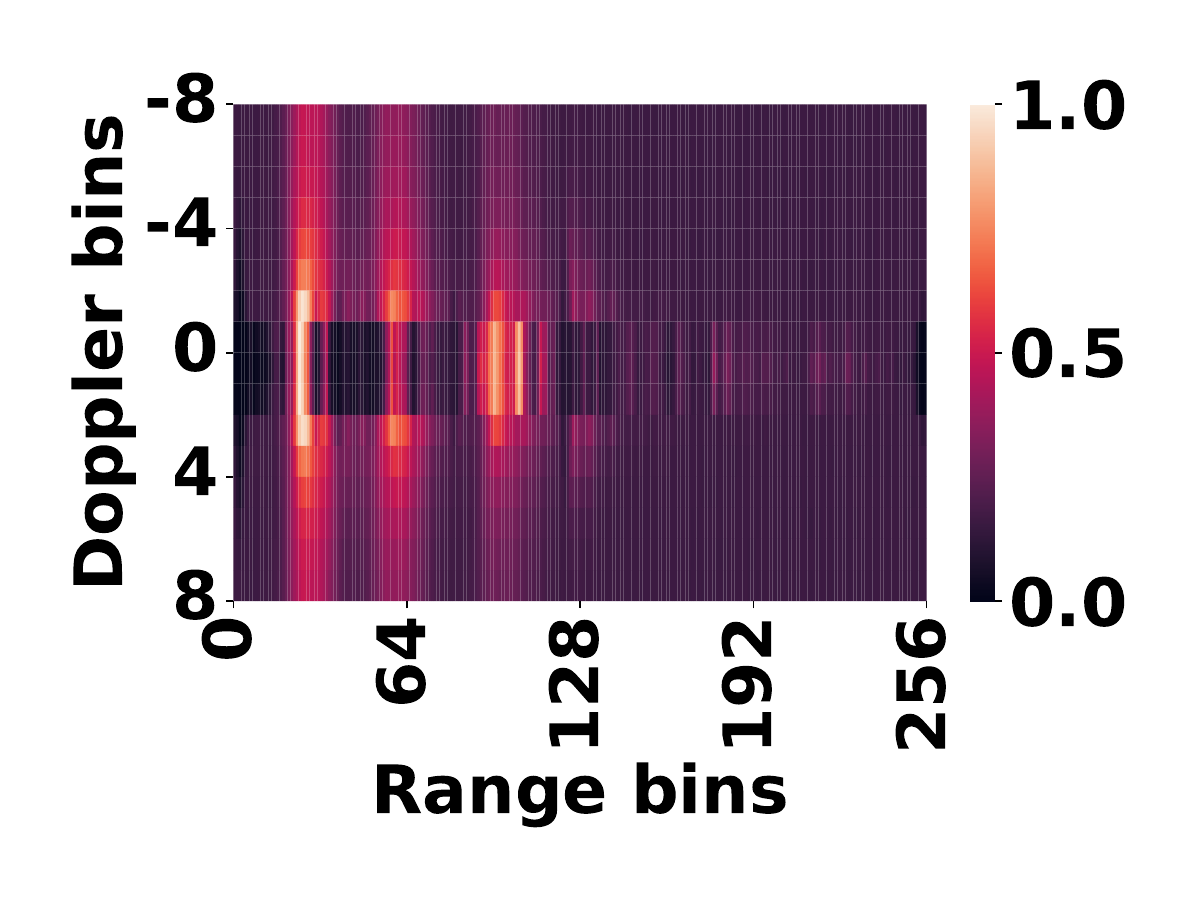}
    }
    \subfloat[Vacuum \\Clean]{
    \includegraphics[trim = {22mm 0 0 7mm}, width=\wide\columnwidth]{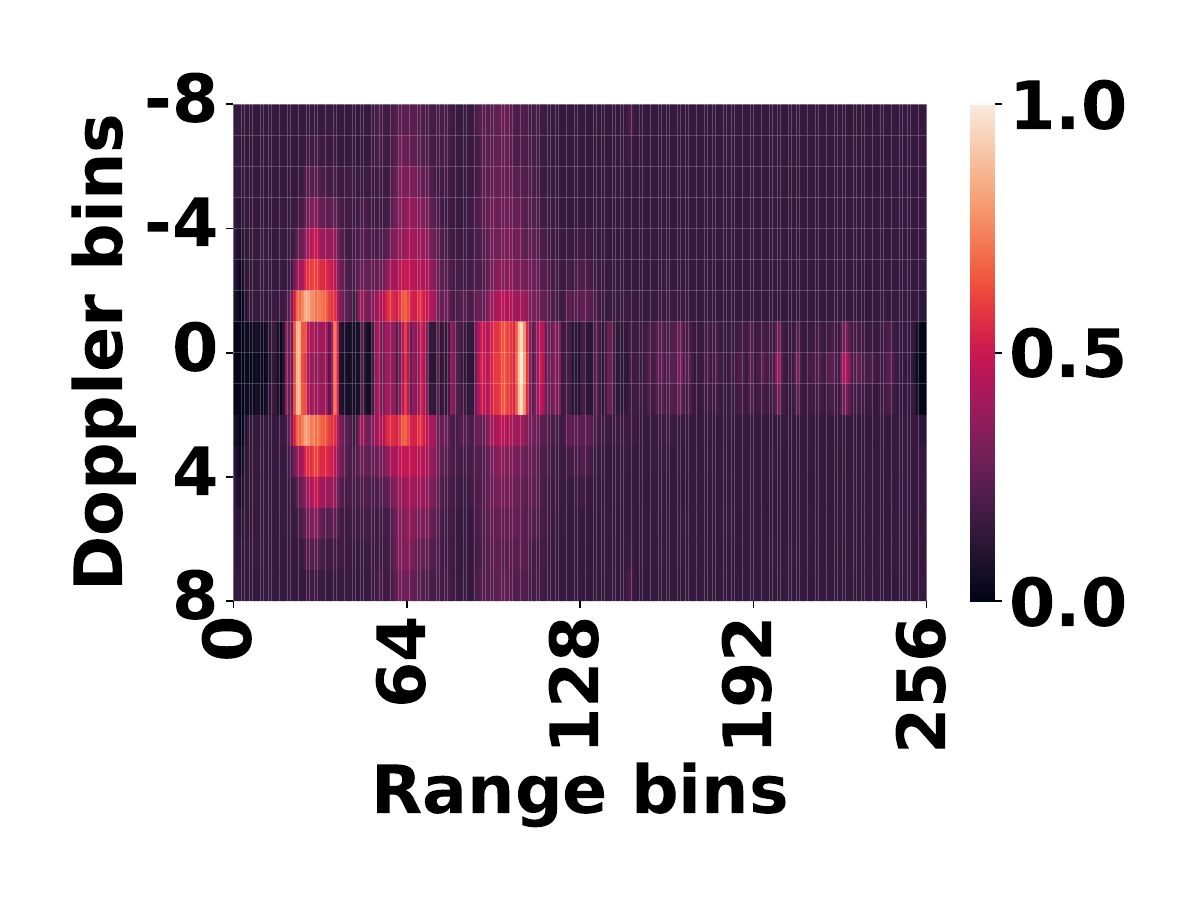}}
    \subfloat[Running]{
    \includegraphics[trim = {22mm 0 0mm 7mm}, clip, width=\wide\columnwidth]{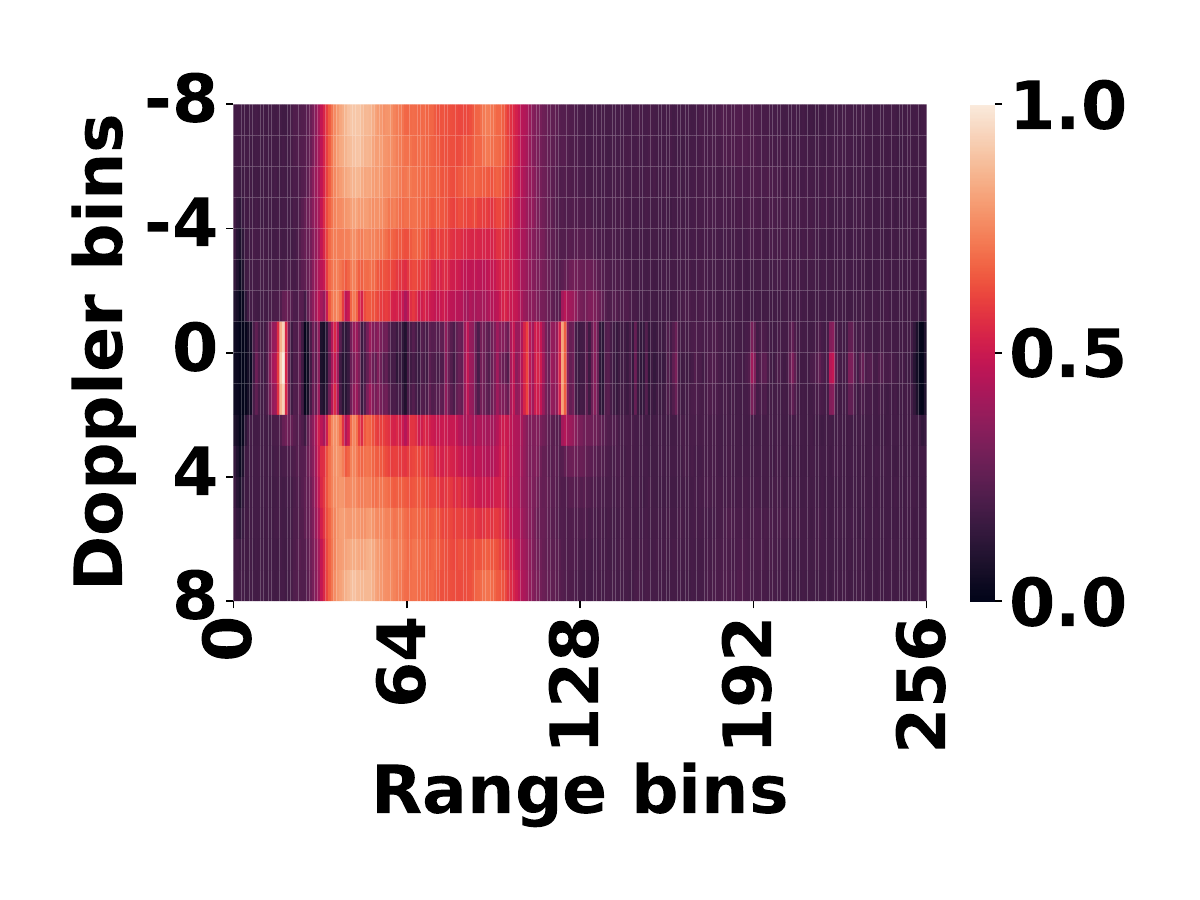}
    }\\
    \vspace{-5pt}
        \subfloat[Phone \\Type]{
    \includegraphics[trim={22mm 0 0 7mm},width=\wide\columnwidth]{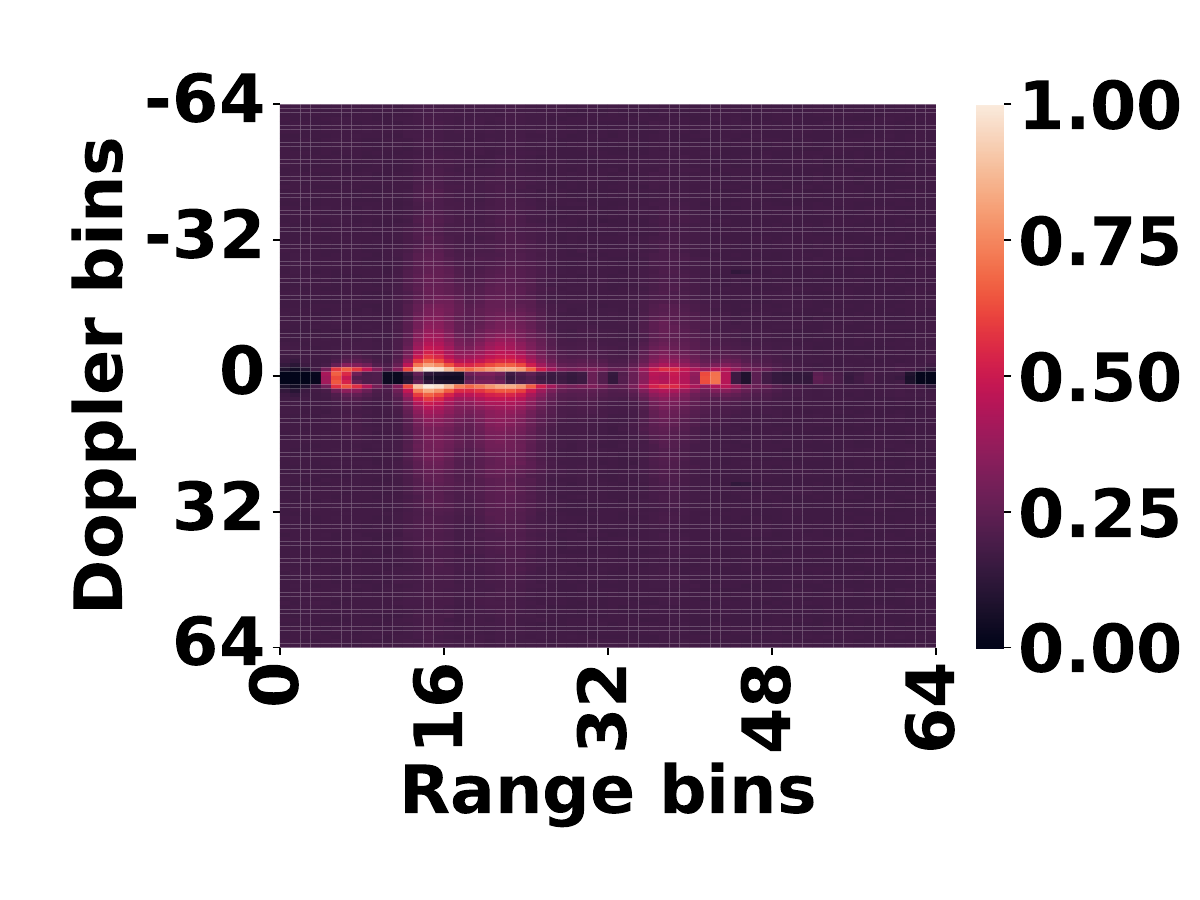}
    }
    \subfloat[Laptop \\Type]{
    \includegraphics[trim={22mm 0 0 7mm},width=\wide\columnwidth]{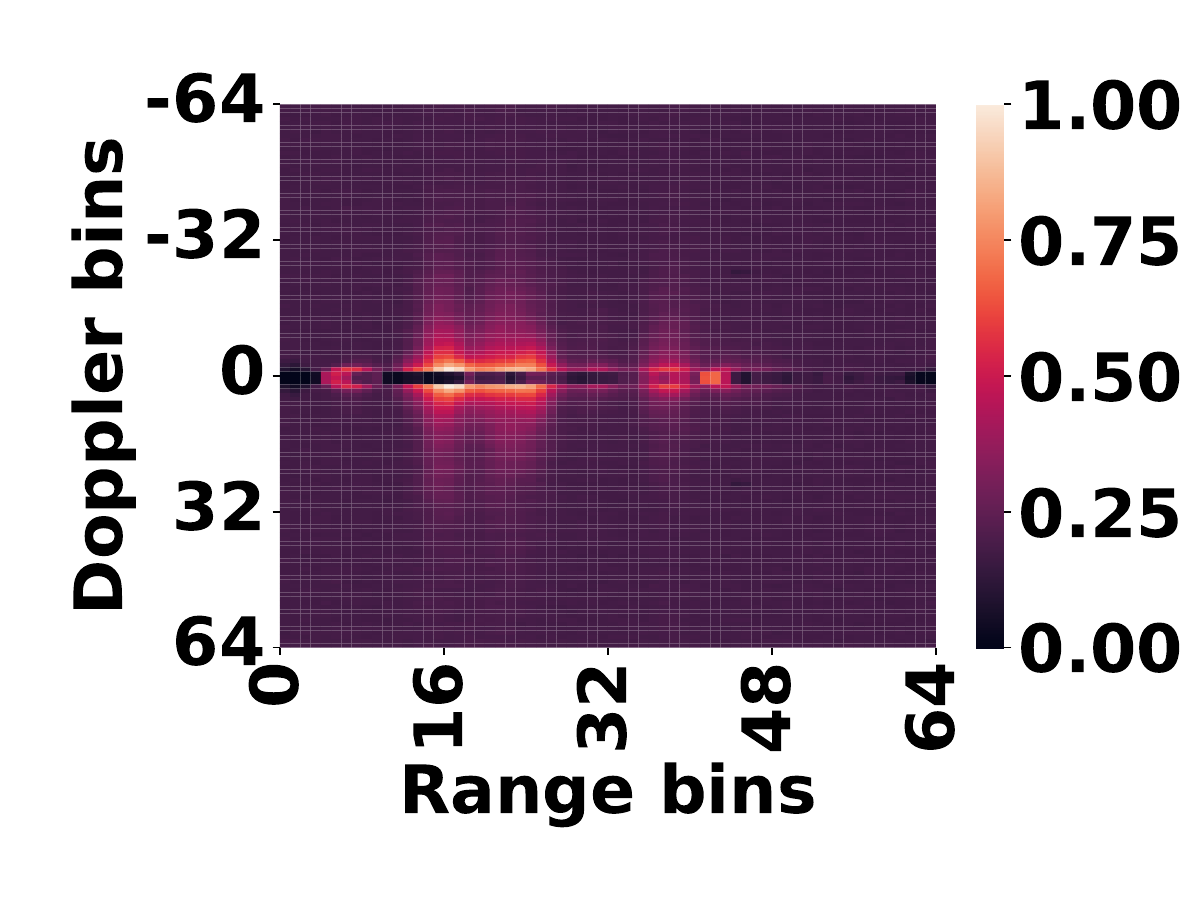}
    }
    \subfloat[Sitting]{
    \includegraphics[trim={22mm 0 0 7mm},width=\wide\columnwidth]{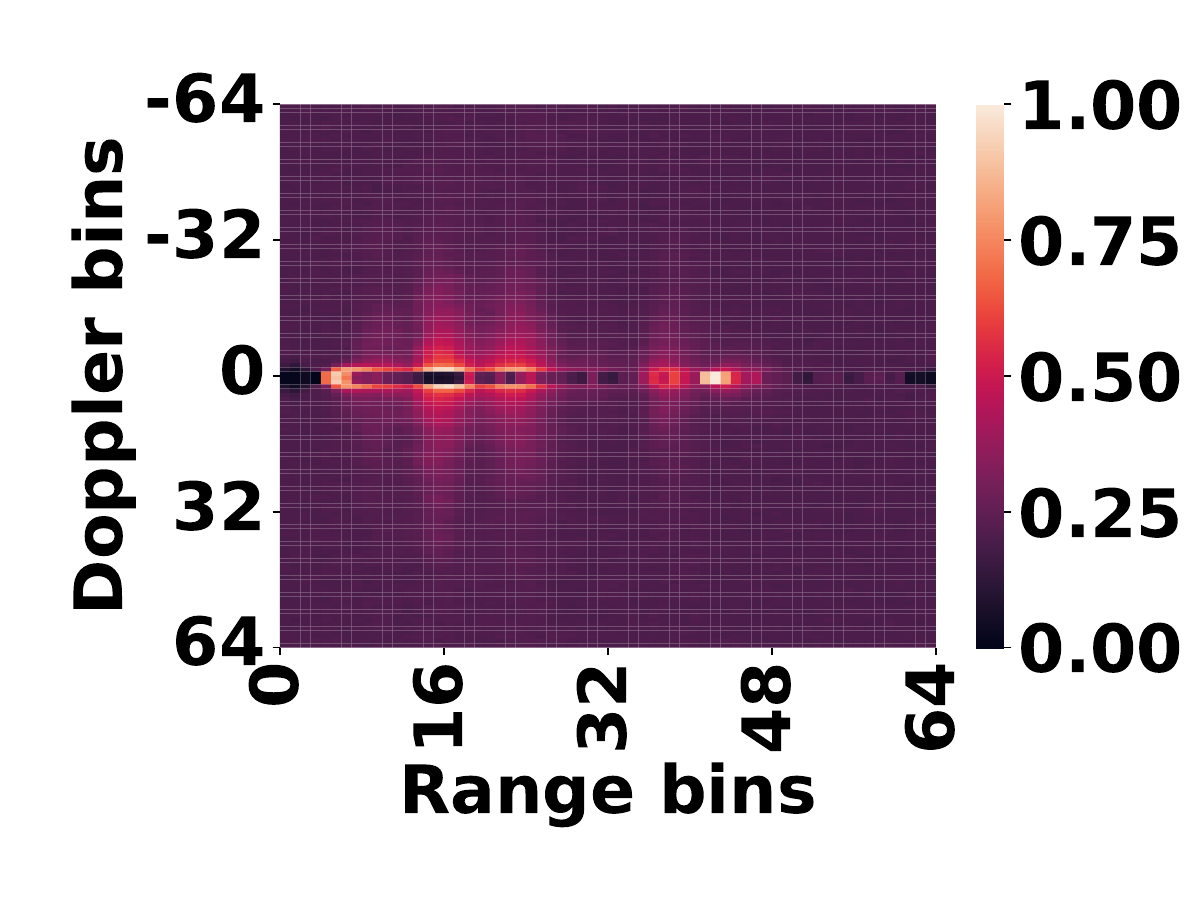}
    }
        \subfloat[Eating]{
    \includegraphics[trim={22mm 0 0 7mm}, clip, width=\wide\columnwidth]{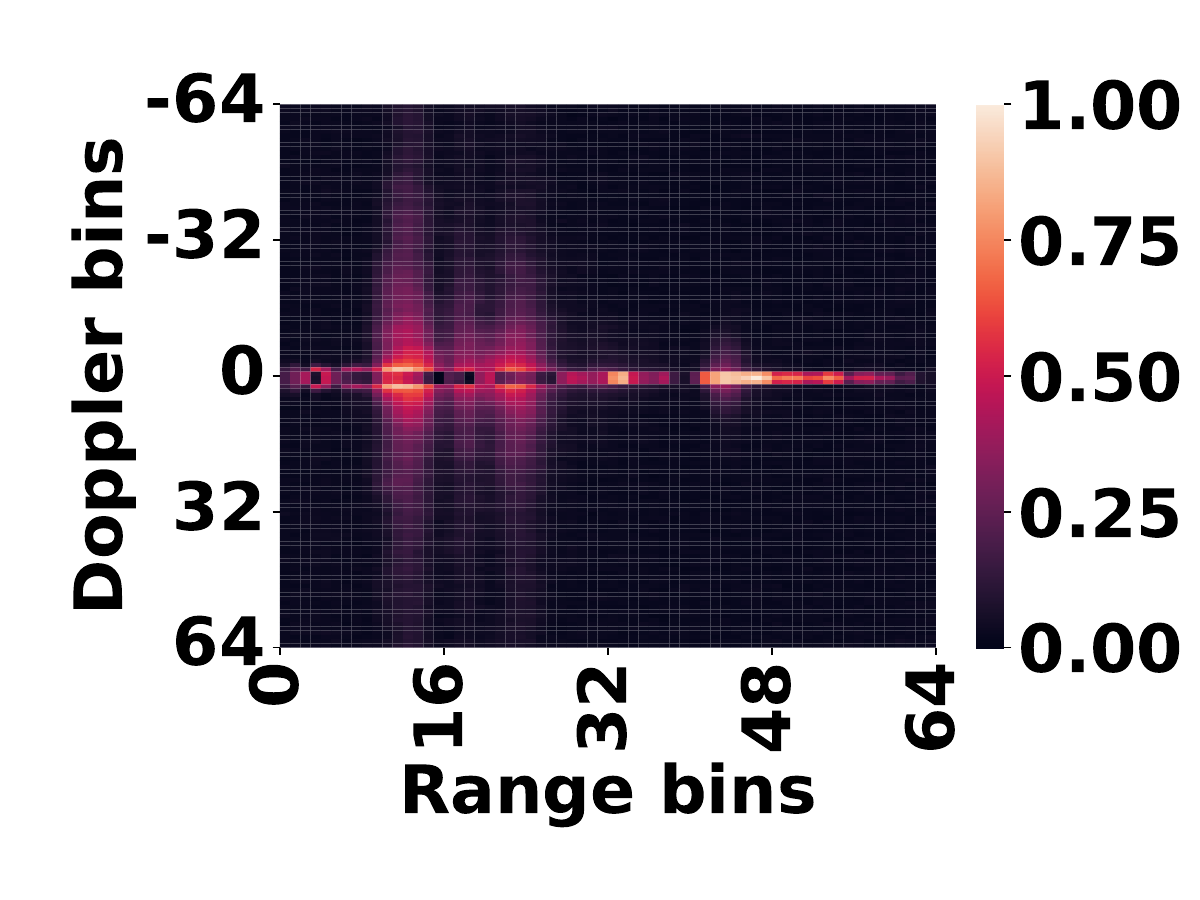}
    }
    \subfloat[Phone \\Talk]{
    \includegraphics[trim={22mm 0 0 7mm},width=\wide\columnwidth]{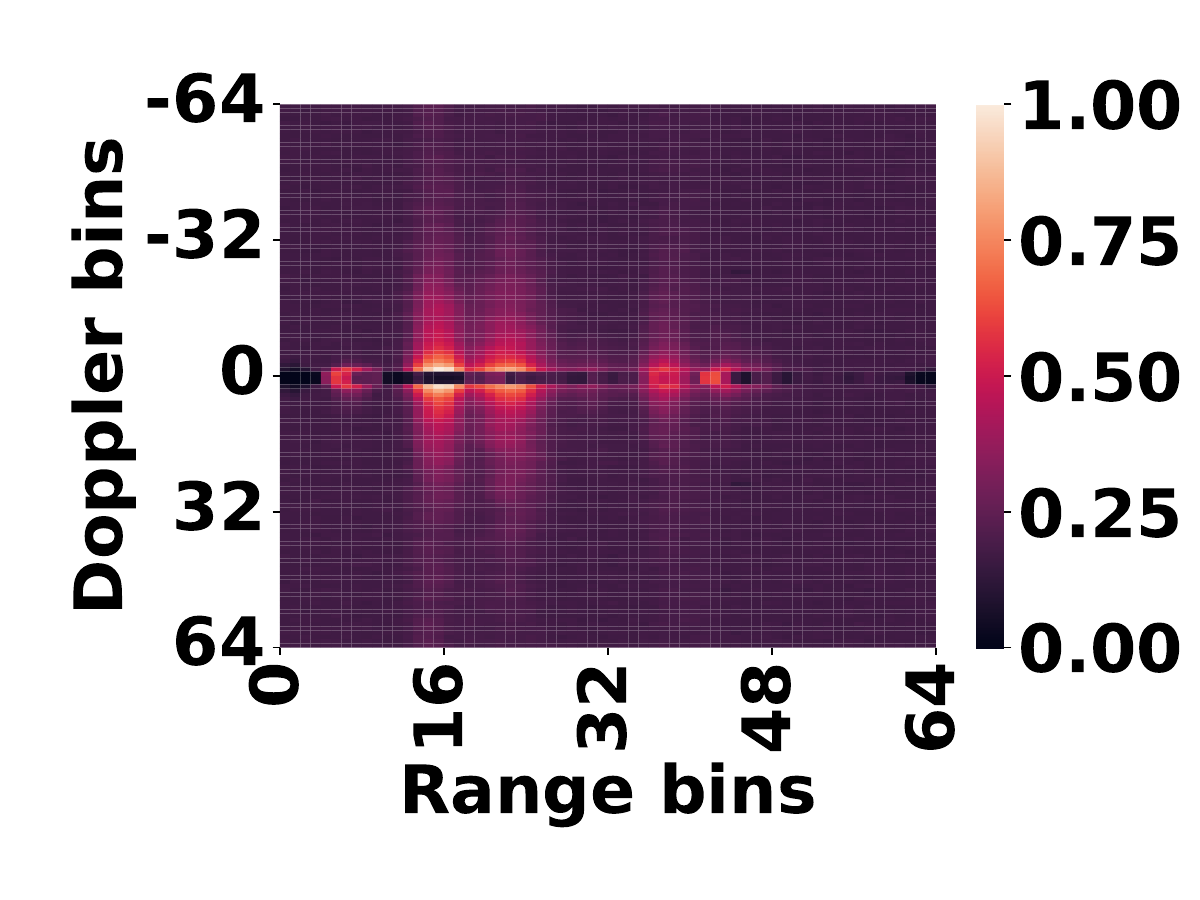}
    }\\
    \vspace{-5pt}
    \subfloat[Play \\Guitar]{
    \includegraphics[trim = {22mm 0 0 7mm}, width=\wide\columnwidth]{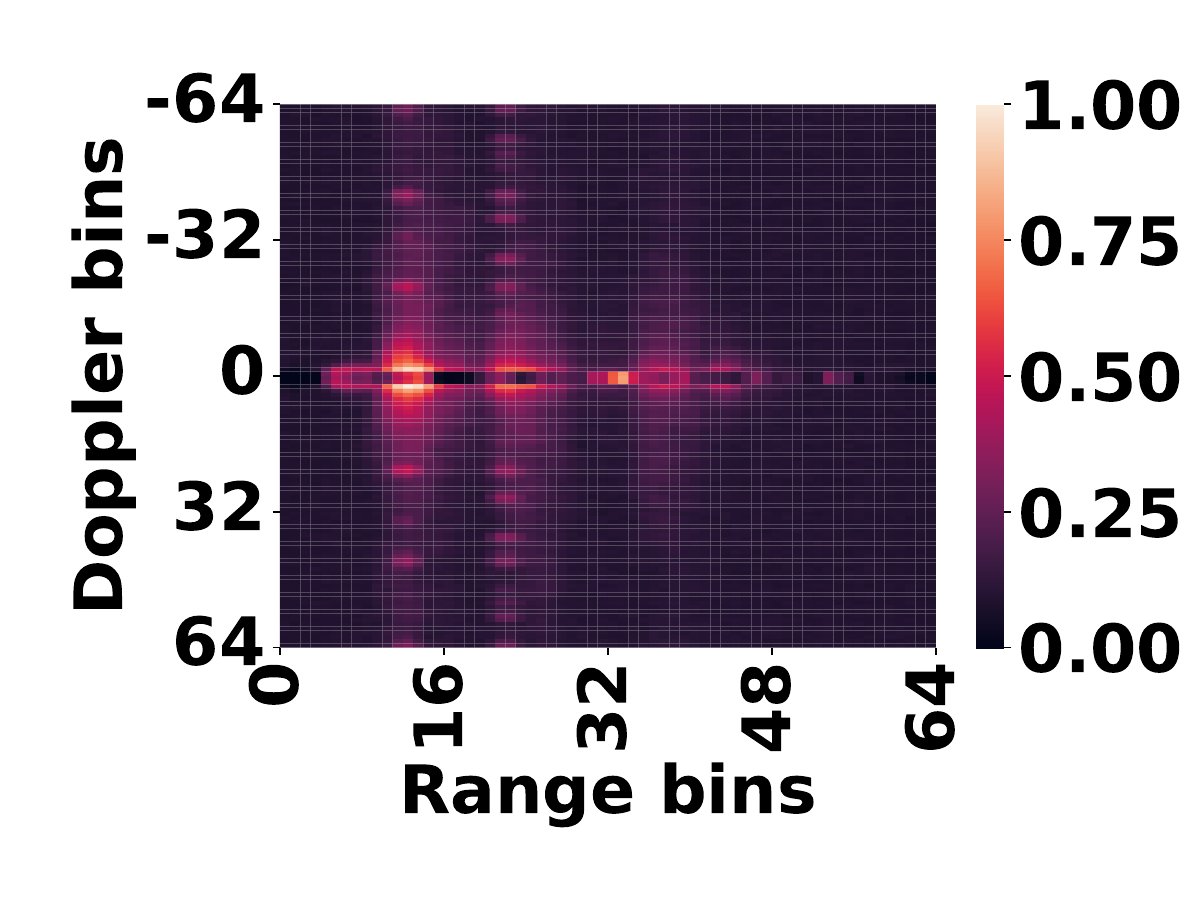}
    }
    \subfloat[Brushing]{
    \includegraphics[trim = {22mm 0 0 7mm}, clip, width=\wide\columnwidth]{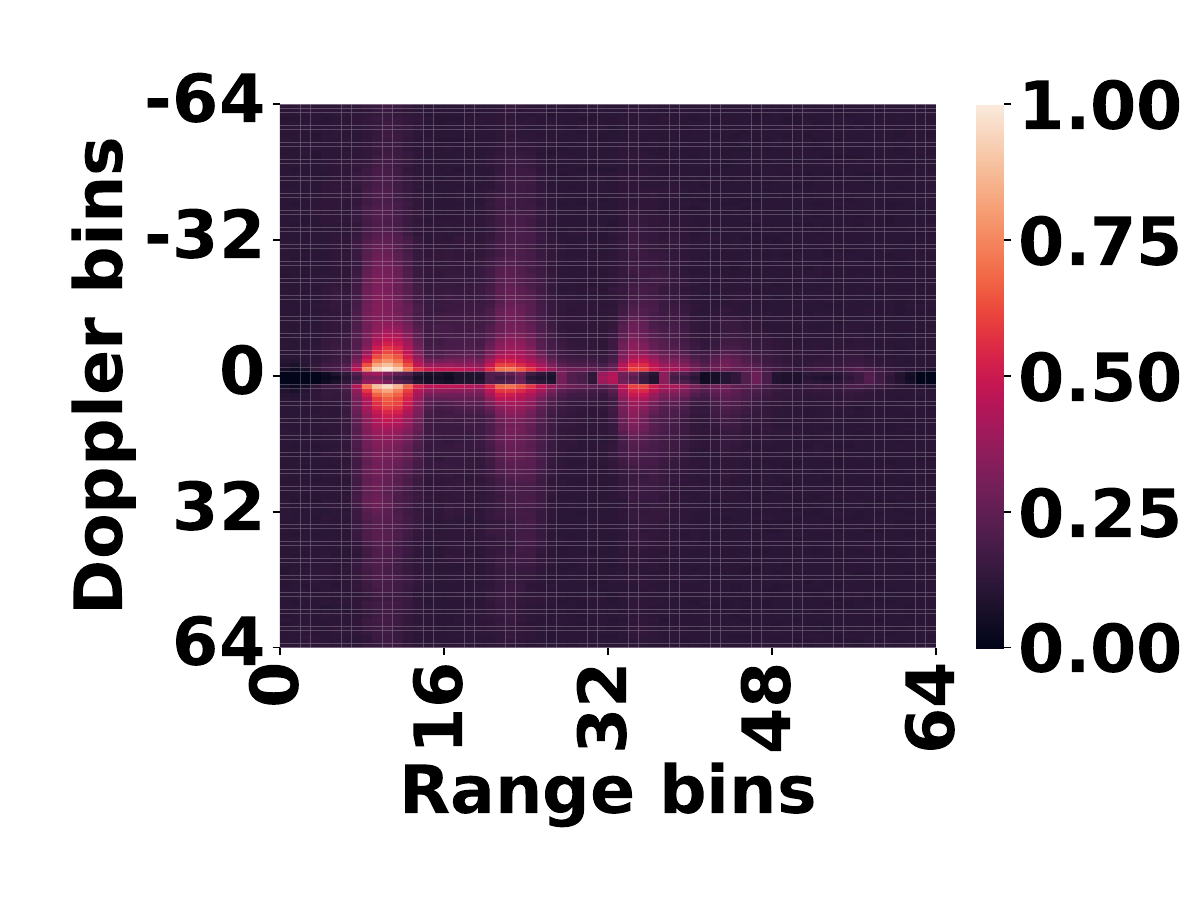}
    }
    \subfloat[Combing]{
    \includegraphics[trim = {22mm 0 0 7mm}, clip, width=\wide\columnwidth]{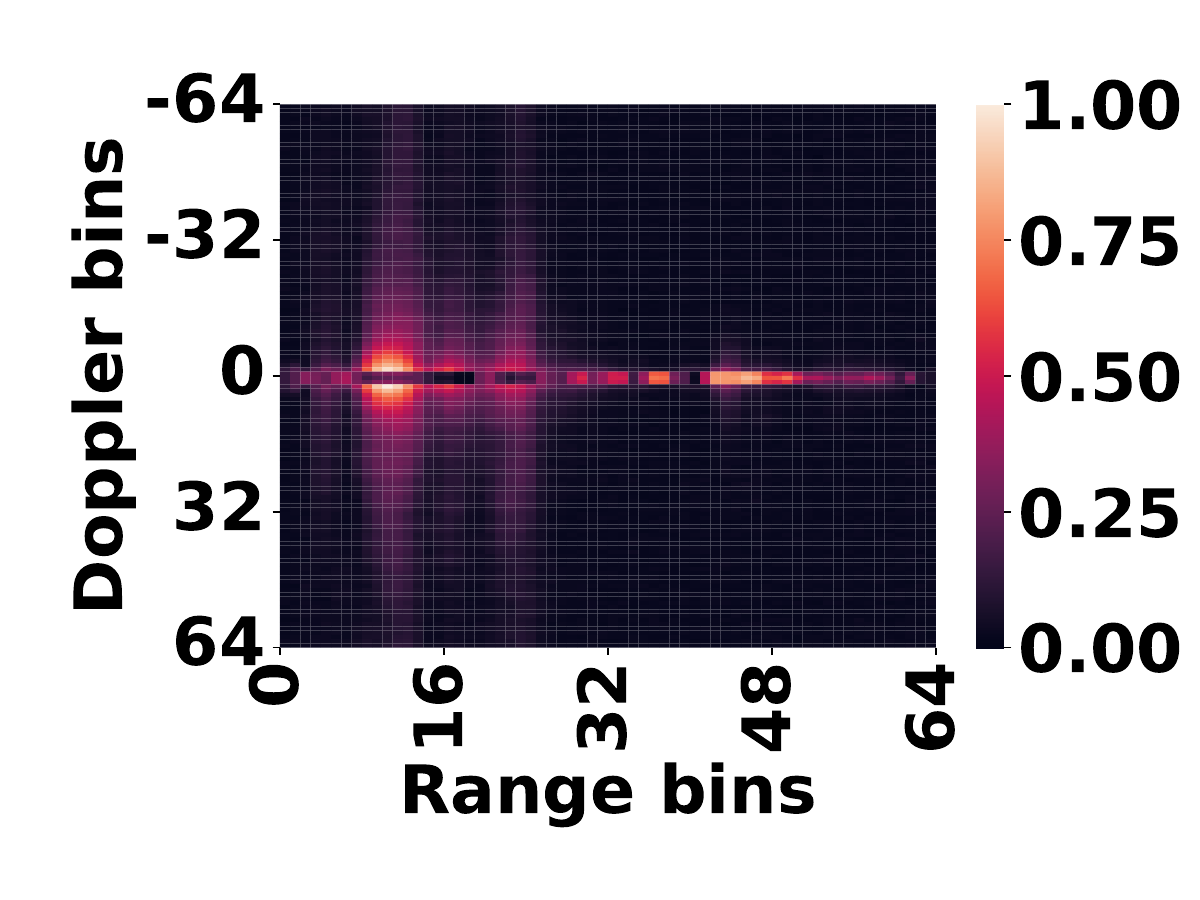}
    }
    \subfloat[Drinking]{
    \includegraphics[trim = {22mm 0 0 7mm}, width=\wide\columnwidth]{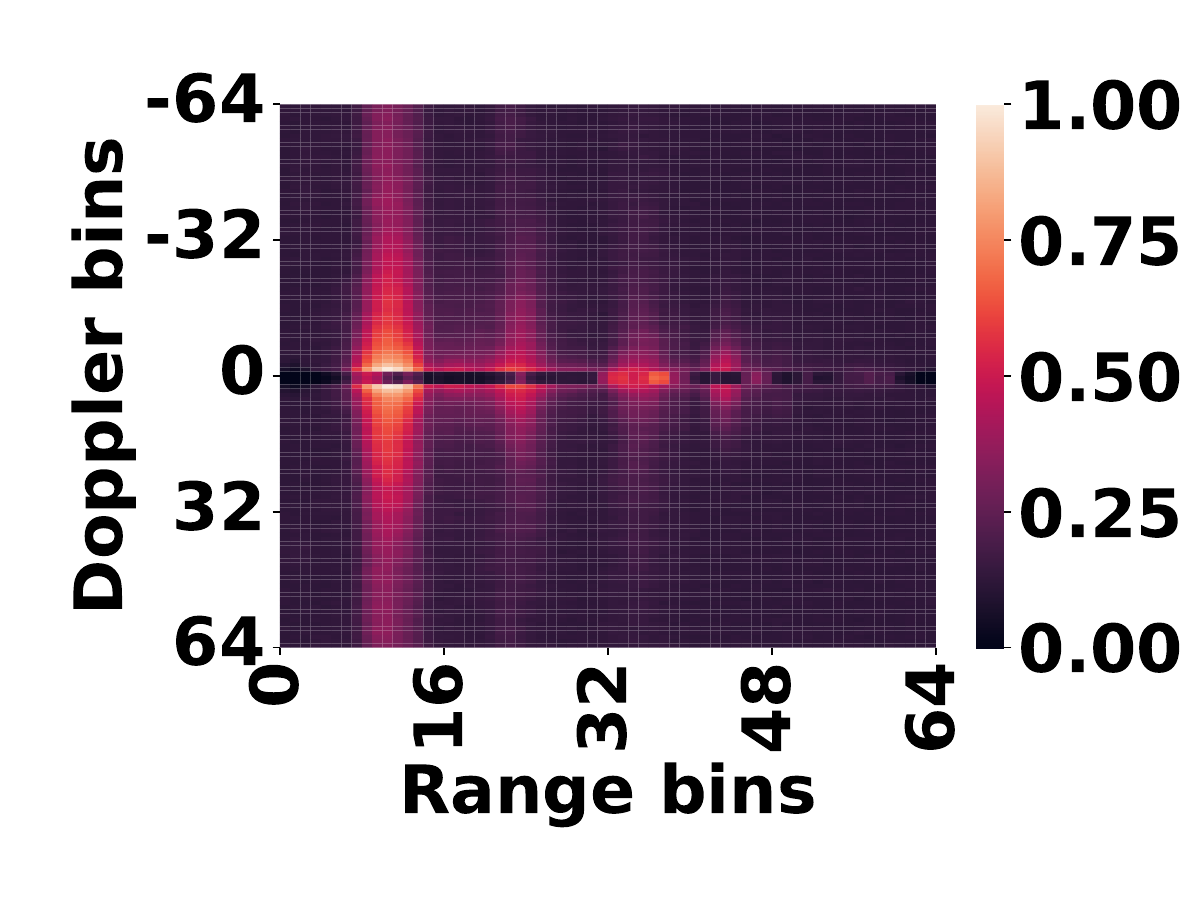}
    }
 \caption{Standard deviation (std) in the range-doppler heatmaps captured during the entire activity duration. (a)-(j): Macro activities with low doppler resolution, (k)-(s): Micro activities with high doppler resolution. Activities having similar body movements have similar patterns, but the difference can be captured in the temporal domain.}
\label{fig:dopplerpatterns}
 \end{figure*}

\begin{figure*}[t]
    \centering
    \includegraphics[width=0.62\textwidth]{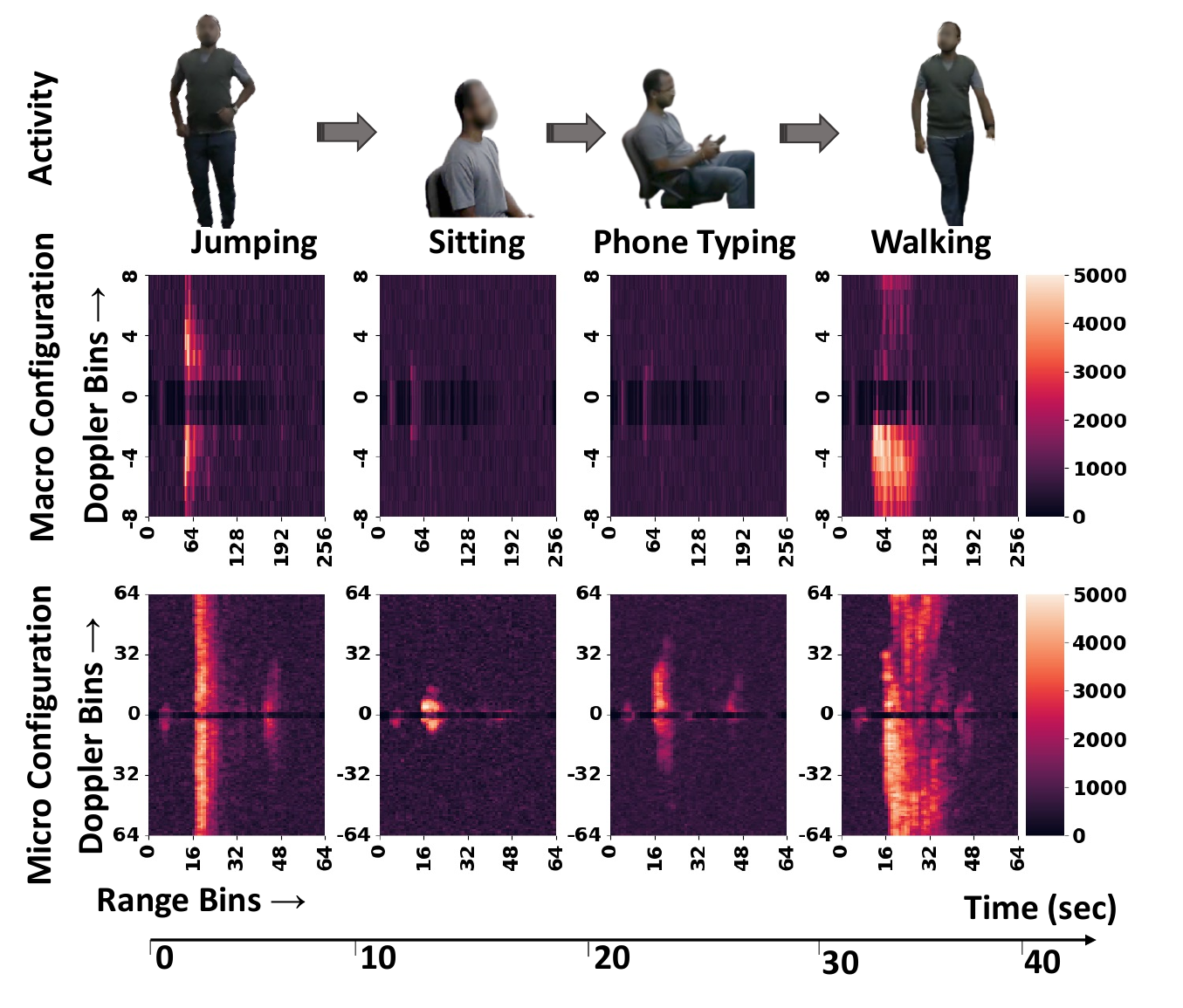}
    \caption{Range-doppler signatures (standard deviation for individual activity windows) over time.}
    \label{fig:conf_motivation}
\end{figure*}

\subsection{Observations from the heatmaps}
We perform a set of observations that establish the effectiveness of the range-doppler signatures for capturing diverse activity scenarios. We detail these observations below.
\subsubsection{Feasibility study for range-doppler}
\figurename~\ref{fig:dopplerpatterns} shows the standard deviation in the range-doppler heatmaps captured during the activity. Notably, standard deviation technique removes static powers (-3 to 2 doppler bins) in the heatmap; thus, we see a low-power value in these doppler bins. We observe that each activity has different signatures captured by the range-doppler heatmap. Although the plotted standard deviation looks similar for some activity pairs with similar body movements (like walking/running, jumping/lunges), there are temporal changes in the heatmaps; for example, ``running'' induces a faster change than ``walking''. Thus, combining the observations from range, doppler, and time, we can get different signatures. Notably, the macro activities have more robust patterns due to the magnitude of movement involved. Even though the micro activities have relatively weaker signatures, they can be distinctively captured with a \textit{higher doppler resolution} ($-64$ to $+64$ doppler bins, in contrast to $-8$ to $+8$ doppler bins used for macro activities).

\subsubsection{Impact of radar configurations on determining a subject's activity}
To understand how the radar configuration affects the patterns in the activity signatures, we ask one subject to switch his activity from jumping to two micro activities, namely, sitting in a chair and phone typing, and finally, walking out of the room. The subject is asked to repeat the pattern twice to collect the corresponding range-doppler data under \textit{low and high doppler resolution}. From the std in the heatmap across the entire activity time axis \figurename~\ref{fig:conf_motivation}, it is evident that low doppler resolution is adequate for capturing macro activities like walking and jumping. Still, typing and sitting does not have any significant signatures. On changing the radar configuration to \textit{high doppler resolution}, we observe that micro activities like typing and sitting have better visibility. However, with this, the macro activities (walking, jumping) generate noisy data due to the higher resolution. Therefore, \textit{different doppler resolutions} are crucial to capture the signatures corresponding to different activity types.  

\subsubsection{Multi-user Activity}
\begin{figure*}
    \centering
    \subfloat[]
    {\includegraphics[width=0.38\textwidth]{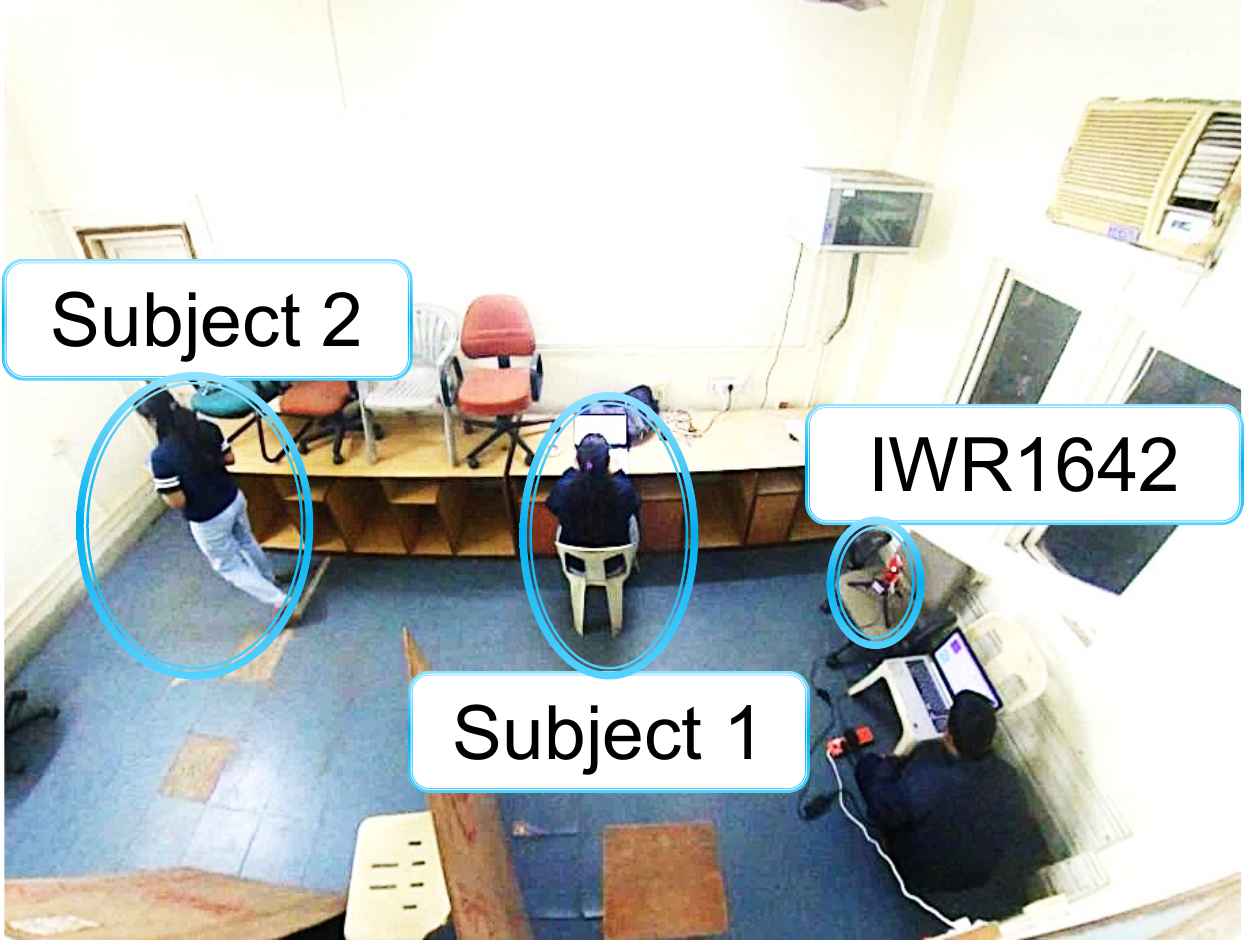}\label{fig:static_scenario1}}\hspace{10pt}
     \subfloat[]{\includegraphics[width=0.42\textwidth]{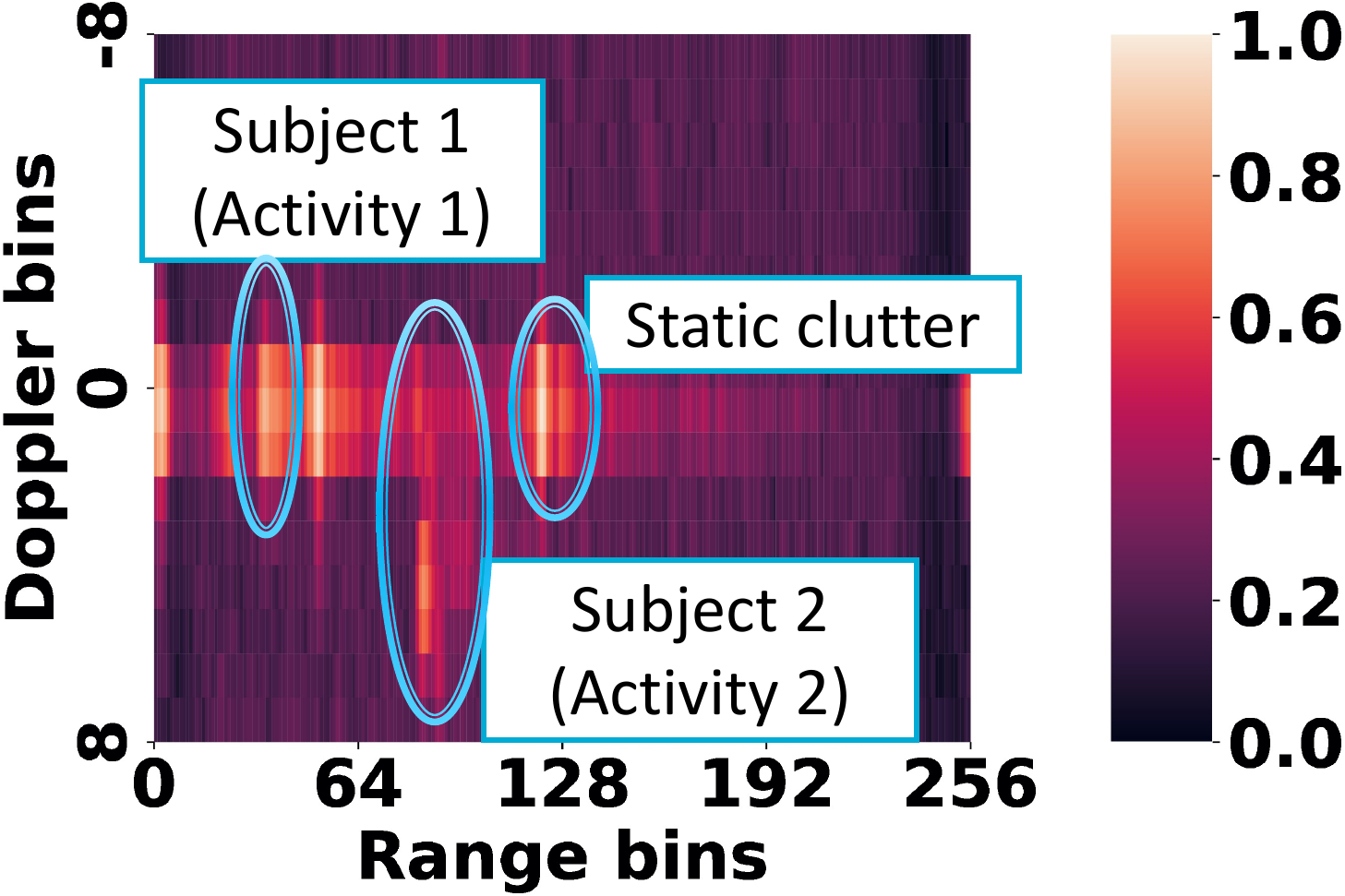}\label{fig:range_doppler_activity}}
     \caption{Range Doppler signature of two subjects performing different activities simultaneously}\label{fig:mul_act}
\end{figure*}

Now, we elaborate on how to identify individual activity signatures when multiple subjects are of concern. It is worth noting that the captured range-doppler heatmaps often contain \textit{static clutters}. Static clutters are any objects (walls, furniture, etc.) that are stationary but can reflect the mmWave signal and generate unwanted signatures in the range-doppler data. To elaborate on this, let us consider a scenario with two subjects, \textit{Subject 1 sitting and Subject 2 walking}, inside the room, as shown in \figurename~\ref{fig:static_scenario1}. The room also contains multiple static clutters, such as wooden sheets and walls. From the range-doppler heatmap as shown in, \figurename~\ref{fig:range_doppler_activity} we observe multiple peaks at the range bins corresponding to the subjects and the static clutter. Here, we observe that the static clutter produces a higher magnitude along \textit{the zero Doppler axis}, thereby signifying zero or no movement. On the other hand, the dynamic movements of the subjects are positioned across \textit{non-zero doppler bins.} A major takeaway from the range-doppler heatmap is that static clutters are easily identifiable by their zero-doppler signatures.
\subsection{Benchmark Evaluation}

The 2D range-Doppler heatmaps exhibit unique spatial patterns for each activity. However, the variations within each activity are confined to a limited range of bin regions, resulting in predominantly sparse data. To address this, we utilize a 2D convolutional Neural Network (2D-CNN) architecture to extract key features from these sparse heatmaps. The 2D convolution operation leverages the dependency of neighboring spatial values, and by concatenating multiple past frames in a sliding window fashion, we capture the temporal relationships over the past $t$ frames.

For macro and micro activity feature extraction, we employ four and three 2D convolutional layers respectively, both with the `same' padding and ReLU activation. Following these layers, a global average pooling layer is used to extract the average spatial activation across the entire feature map. The feature extraction model is trained separately for macro and micro activities using $90\%$ of the entire dataset. The remaining $10\%$ of the dataset is used as the test set, where the final feature embeddings are obtained from the global average pooling layer. These embeddings are then visualized in a reduced 2-dimensional space using t-SNE.

\begin{figure*}
    \centering
    \subfloat[]
    {\includegraphics[width=0.32\textwidth]{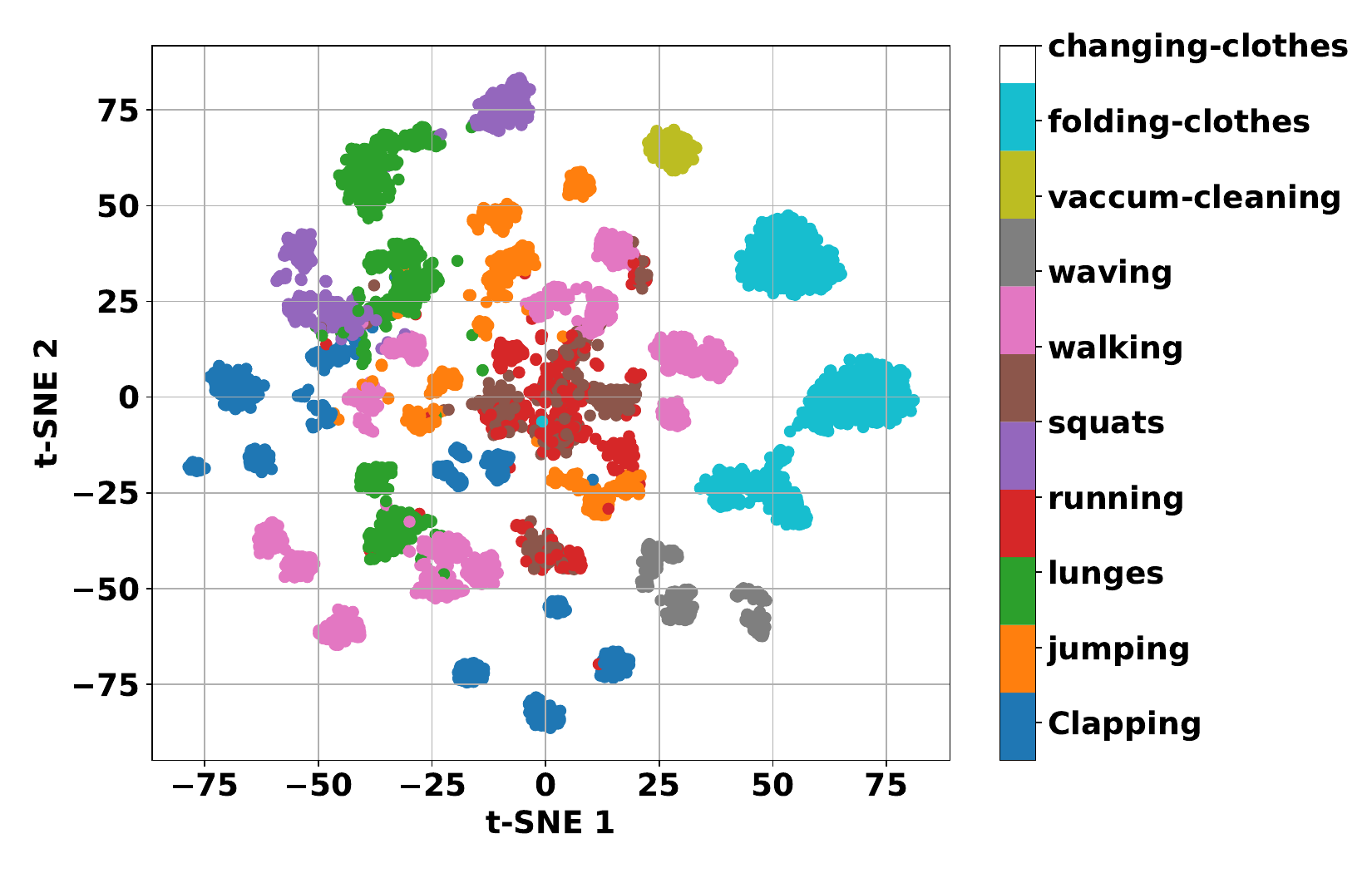}\label{fig:macro_tsne}}
     \subfloat[]
     {\includegraphics[width=0.32\textwidth]{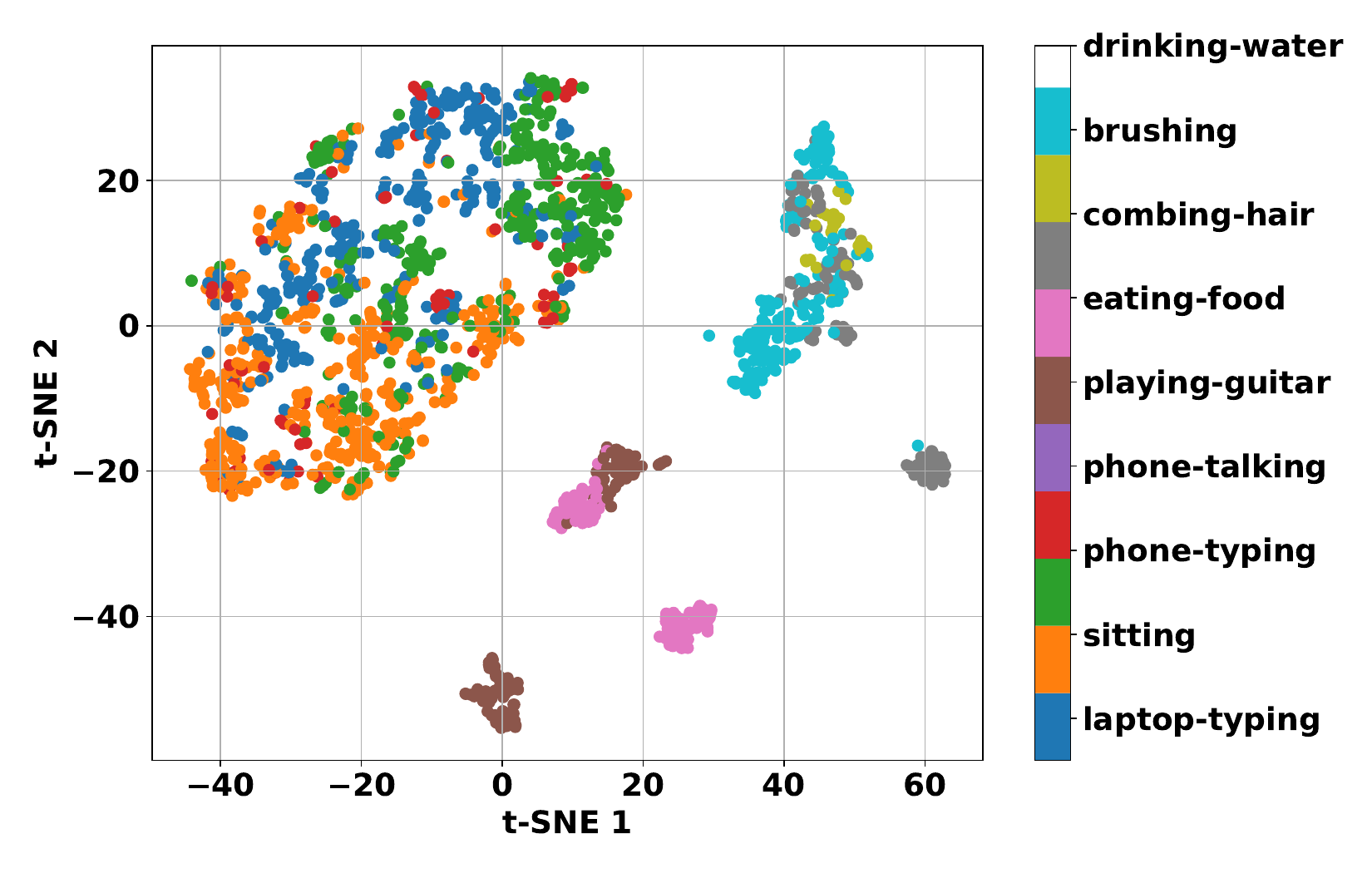}\label{fig:micro_tsne}}
      \subfloat[]{\includegraphics[width=0.28\textwidth]{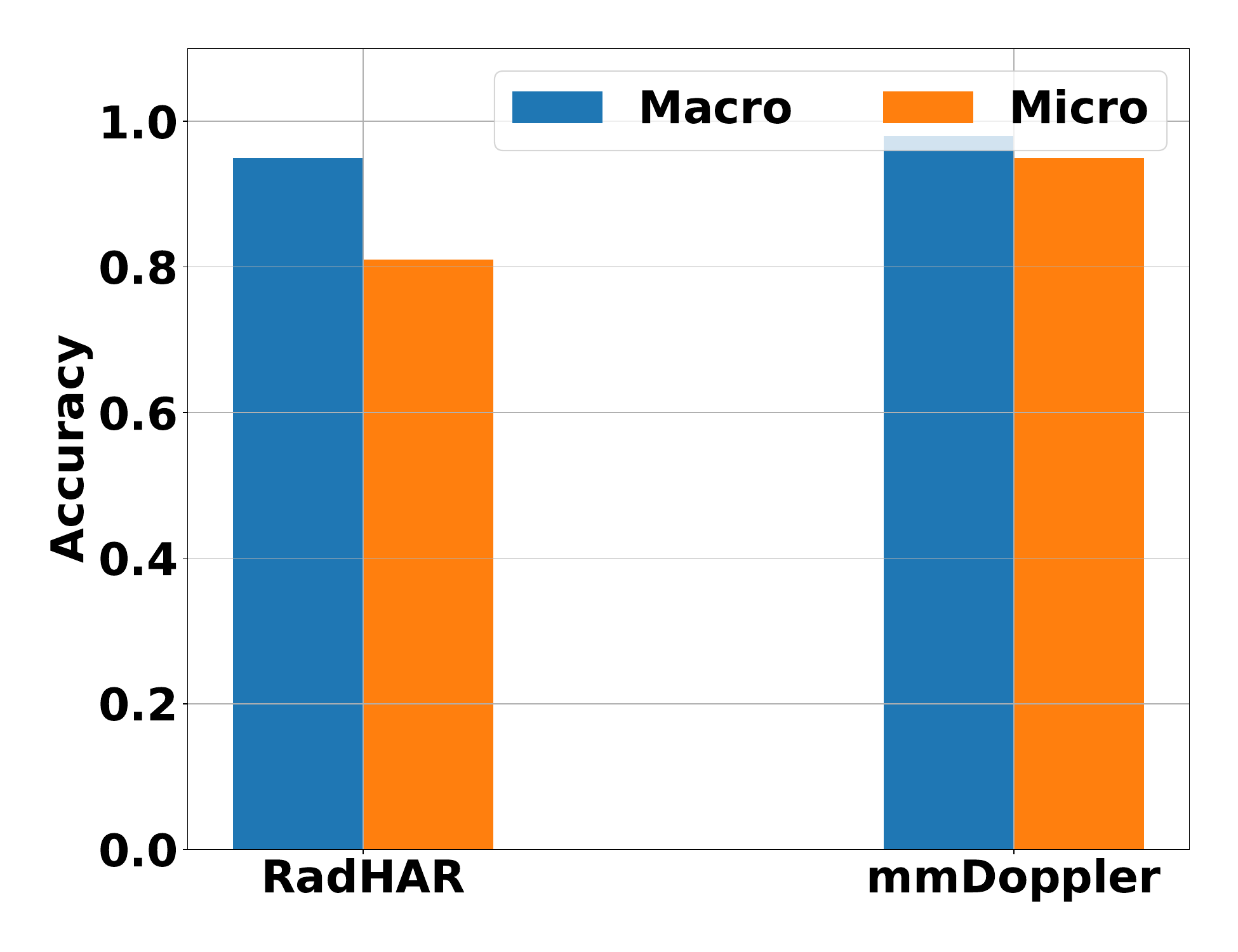}\label{fig:comp}}
     \caption{(a) \& (b) t-SNE Distribution of Extracted Feature Embeddings for Macro and Micro Activities, (c) Comparison with pointcloud-based existing HAR classifier (RadHAR) with range-doppler based classifier (\ourdataset).}
\end{figure*}

As illustrated in \figurename~\ref{fig:macro_tsne} and \figurename~\ref{fig:micro_tsne}, the feature embeddings from the test dataset form distinct clusters for different activities. This clustering underscores the dataset's capability to effectively capture and differentiate between various human activity patterns. Additionally, when we compare this proposed classifier taking range-doppler heatmaps, it provides an accuracy of 98\% and 95\% for macro and micro activities, respectively. Whereas, the collected pointcloud data for the same set of activities when studied by a pointcloud-based HAR classifier, RadHAR~\cite{singh2019radhar}, we observe, although the classifier is performing well with for macro activities, it fails in the case of micro activities as pointcloud cannot capture human movements at a higher granularity.  The results are shown in \figurename~\ref{fig:comp} 

\section{Conclusion}
\ourdataset represents a significant advancement in the field of human activity recognition (HAR), offering a comprehensive and versatile resource for researchers and developers. By integrating range-Doppler heatmaps and point cloud data by employing adaptive doppler resolution, \ourdataset bridges the gap between macro and micro-scale activity recognition. This allows for the precise detection of both large and small movements, capturing the nuances of human behavior in diverse indoor settings. The dataset includes 19 different activities performed by multiple subjects, ensuring a rich variety of real-world scenarios and interactions. By making \ourdataset publicly available, we hope to foster further research and innovation in HAR, enabling the development of robust, accurate, and versatile systems for healthcare, security, and other uses.
\begin{acronym}
	\acro{mmWave}{Millimeter Wave}
	\acro{RF}{Radio Frequency}
	\acro{4G}{4$^\text{th}$ Generation}
	\acro{5G}{5$^\text{th}$ Generation}
	\acro{COTS}{Commercial off-the-shelf}
	\acro{FMCW}{Frequency Modulated Continuous Wave}
	\acro{ADC}{Analog to Digital Conversion}
	\acro{DBSCAN}{Density-Based Spatial Clustering of Applications with Noise}
	\acro{EKF}{Extended Kalman Filter}
	\acro{RKF}{Recursive Kalman Filter}
	\acro{LoS}{Line of Sight}
	\acro{NLoS}{Non Line of Sight}
	\acro{FPS}{Frames Per Second}

\bibliographystyle{abbrv}
\bibliography{refs.bib}
\newpage
\appendix
\section{Appendix for \ourdataset}
\subsection{Preliminaries}\label{sec:prelims}
Before diving into the dataset\footnote{\url{https://github.com/arghasen10/mmdoppler} (Access: \today)} documentation, we first discuss the preliminaries of mmWave sensing to familiarize with the features captured in the dataset, such as pointclouds and range-doppler heatmaps. The primary working principle of Commercial Off-The-Shelf (COTS) mmWave radars revolves around Frequency-Modulated Continuous Wave (FMCW) technology~\cite{rao2020introduction}. These radars emit continuous frequency chirps and perform a process called \textit{dechirping} by mixing the transmitted signal (TX) with the reflected signal (RX) from objects. This creates an \textit{Intermediate Frequency} (IF) signal. From this IF signal, we can derive (1) a \textit{Pointcloud}, which is a discrete set of points representing the detected objects~\cite{schumann2018semantic}, and (2) the \textit{Range-doppler heatmaps}, which are 2D image representations where the abscissa represents the range (location of human subjects), and the ordinate indicates the doppler speed at which subjects are moving. These heatmaps combine range and doppler  information from radar signals, offering a detailed view of motion dynamics over time.

\subsubsection{Range-doppler Estimation}\label{subsec
}
The distance between an object and the radar is determined by measuring the frequency difference between the reflected and transmitted signals~\cite{iovescu2020fundamentals}. This frequency difference, known as the \textit{beat frequency} ($f_b$), occurs after a Round Trip Time (RTT) of $\tau$. As shown in \figurename~\ref{fig:fmcw}, if $T_C$ is the duration of the mmWave chirp across a bandwidth $B$, then the slope of the FMCW chirp is $S = \frac{B}{T_C} = \frac{f_b}{\tau}$.

The RTT delay $\tau$ can be expressed as $\tau = \frac{2d}{c}$, where $d$ is the distance to the detected object and $c$ is the speed of light. Thus, the distance $d$ can be calculated as $d = \frac{c}{2} \cdot \frac{T_C}{B} \cdot f_b$. To determine $f_b$, a Fast Fourier Transform (FFT), known as the \textit{range-FFT}, is performed on the IF signal. This process produces frequency peaks at locations where reflecting objects are present, thereby estimating the range.

\begin{wrapfigure}{r}{5cm}
\centering
\includegraphics[width=0.4\textwidth]{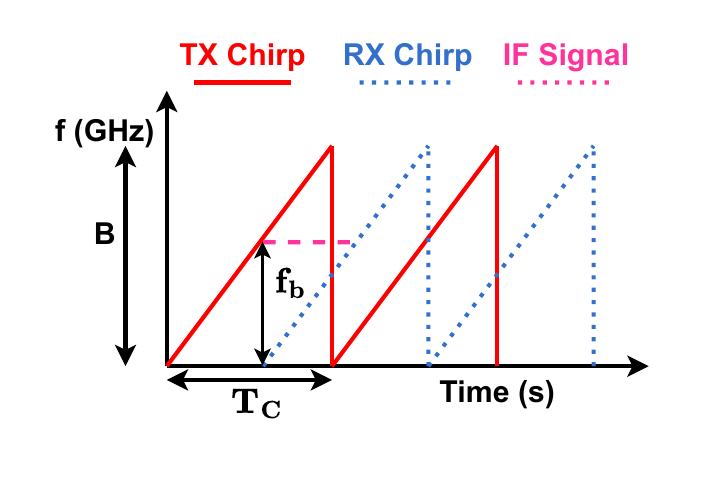}
    \caption{Working Principle of FMCW Radars.}
    \label{fig:fmcw}
\end{wrapfigure}

To measure the velocity or doppler of a moving object, the radar transmits $N$ chirps separated by a duration of $T_C$. If a subject moves at a velocity $v$, the phase difference between two successive RX chirps, corresponding to the motion $vT_C$, is given by $\Delta \phi = \frac{4\pi v T_C}{\lambda}$. A second FFT, called \textit{doppler-FFT}, is performed on these phase differences to determine the object's velocity. This data is captured in a 2D matrix called the \textit{range-doppler} matrix $\mathbb{D}_{D \times R}$, where $D$ and $R$ are the numbers of \textit{doppler bins} and \textit{range bins}, respectively.

Equation \eqref{eq:range_doppler} shows the range-doppler matrix for $R$ range bins and $D$ doppler bins. The range-FFT captures objects at different locations:
\begin{equation}
    \label{eq:range_doppler}
     \mathbb{D} =  \begin{bmatrix}
 \mathbf{P}_{1,1}& \mathbf{P}_{1,2} & \cdots & \mathbf{P}_{1,R}\\ 
 \mathbf{P}_{2,1}& \mathbf{P}_{2,2}& \cdots & \mathbf{P}_{2,R} \\ 
 \vdots & \vdots & \ddots & \vdots\\ 
 \mathbf{P}_{D,1} & \mathbf{P}_{D,2} & \cdots & \mathbf{P}_{D,R} 
\end{bmatrix}_{\text{Doppler Bins} \times \text{Range Bins}}
\end{equation}
For each object, an angular FFT is performed across multiple doppler-FFTs.

\subsection{Pointcloud Estimation}\label{sec
}
The pointcloud is estimated using the standard Constant False Alarm Rate (CFAR) algorithm~\cite{nitzberg1972constant}, which detects peaks in the range-doppler matrix corresponding to detected objects. The pointcloud includes coordinates $(x_i, y_i, z_i)$, doppler variations $(d_i)$, and received power $(p_i)$ of the detected objects. The pointcloud set $S$ for $N$ detected objects is given by $S = \bigcup_{i=1}^N {(x_i, y_i, z_i, d_i, p_i)}$. A sample snapshot of range-doppler heatmaps and pointcloud is shown in \figurename~\ref{fig:mul_acts}.

\begin{figure*}
    \centering
    \subfloat[]
    {\includegraphics[width=0.42\textwidth]{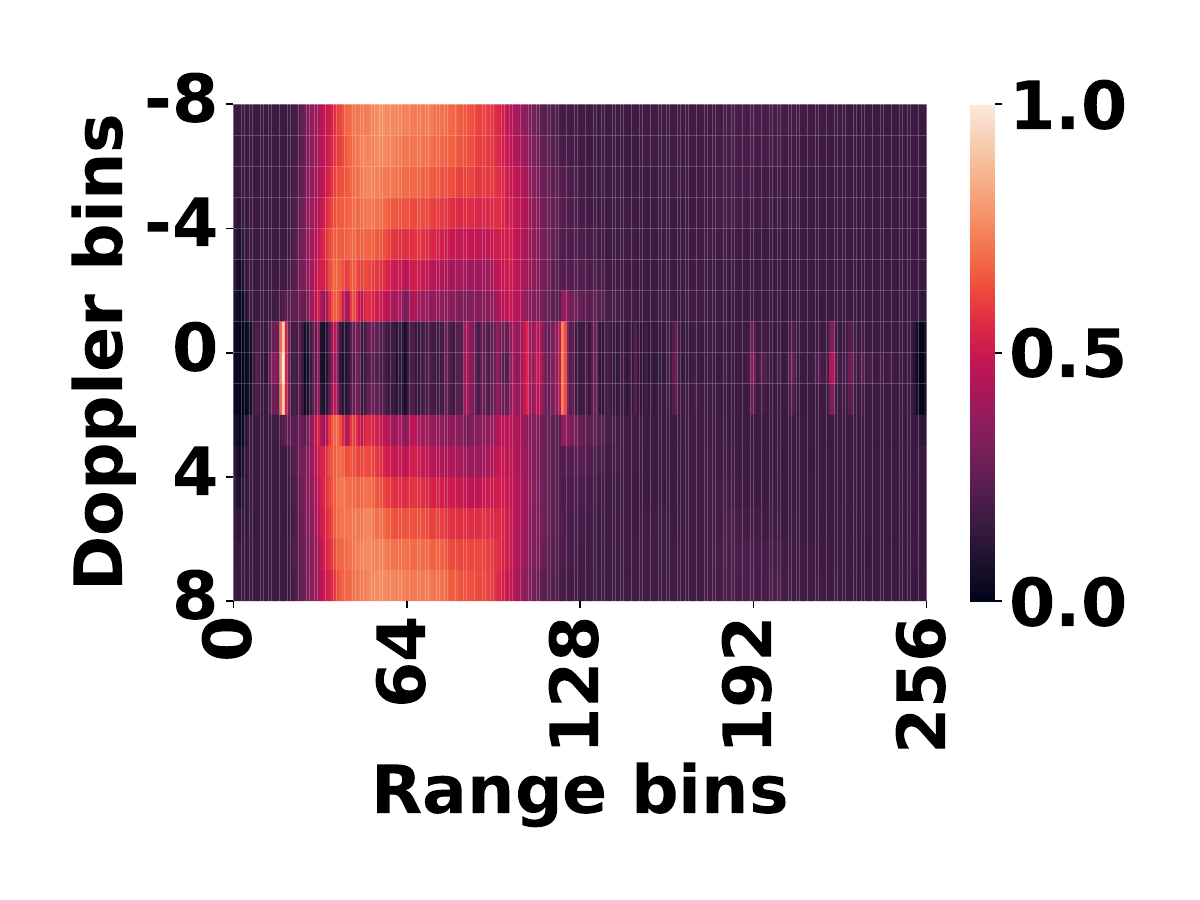}\label{fig:walking}}
     \subfloat[]{\includegraphics[width=0.42\textwidth]{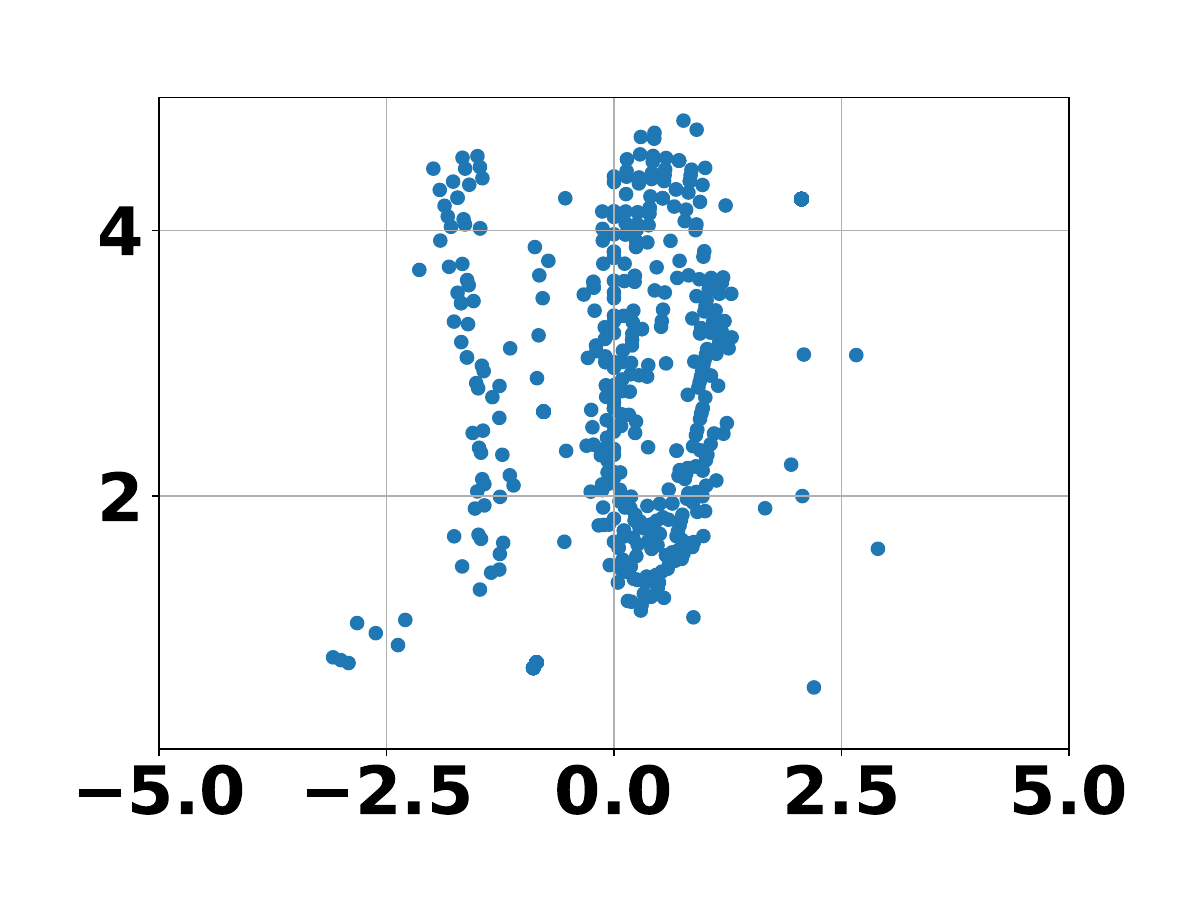}\label{fig:pcd}}
     \caption{Sample Range Doppler and pointcloud signatures for subject performing walking}\label{fig:mul_acts}
\end{figure*}

\subsection{Dataset Documentation}
Our dataset comprises a wide array of 19 different activity classes from Activities of Daily Living (ADLs), Instrumental Activities of Daily Living (IADLs)~\cite{adlsiadls}, and daily indoor exercises: (i) macro activities like walking, running, jumping, clapping, lunges, squats, waving, vacuum cleaning, folding clothes, changing clothes, and (ii) micro activities like laptop-typing, phone-talking, phone-typing, sitting, playing guitar, eating food, combing hair, brushing teeth, and drinking water to ensure comprehensive coverage of daily human behaviors. For macro activities, since higher body movements are involved, a lower doppler resolution can easily capture those activities, while for micro activities, we leverage a higher doppler resolution as it involves subtle body movements. Below, we describe the data collection steps.
\subsubsection{Dataset Collection Steps}
For data collection, \texttt{IWR1642BOOST}~\cite{iwr1642boost} mmWave radar is first connected to the PC via USB cable. One needs to install mmWave Demo Visualizer\footnote{\url{https://dev.ti.com/gallery/view/mmwave/mmWave_Demo_Visualizer/ver/2.1.0/}} from Texas Instruments as documented on our GitHub repository~\footnote{\url{https://github.com/arghasen10/mmDoppler}}. Using this tool, the data gets saved on the local machine in a txt file in JSON format. We kept the saved file nomenclature as \texttt{username\_activity\_iteration}. We have collected the data by asking the subjects to perform the activities at different distances from the radar as well as with different orientations. Along with the raw sensor data collection, we connected a USB camera to the same device to collect video footage of the subject performing several activities. Due to privacy concerns, we have not uploaded the video footage of the subject performing the activities.

\subsubsection{Dataset Labeling}
The raw datasets, stored in \texttt{``*.txt"} format, are first converted into a single \texttt{pandas} DataFrame, with two additional columns: (i) subject name and (ii) activity name, which are derived from the corresponding filename. However, the filename alone does not ensure that the subject performed only that activity throughout the session. For instance, during a jumping activity session, the subject might take short breaks due to fatigue and resume the session after a 10-15 seconds gap.

To refine the coarse-grained activity annotations from the filename, we utilize video recordings from the session to obtain accurate activity annotations. The subjects are instructed to perform the activities at various orientations—left, right, front, and back—relative to the radar's bore-sight angle, as well as at different distances of 2 m, 3 m, and 5 m. These orientations and distances are meticulously annotated based on video recordings to ensure precise data labelling. These annotations, along with timestamps captured from the video, are stored in a \texttt{``.csv"} file. We then merge the raw DataFrame with the annotations from the CSV file to update the activity labels in the dataset.

We leave the annotations blank for periods where the subject is not performing any activities. Consequently, when merging the datasets, these unannotated entries are ignored in the final processed dataset. Thus, we only have the relevant data with accurate activity labels in the processed dataset.

\subsubsection{Feature Space Description}
The collected and processed dataset consists of the following information: (i) datetime, the date and time when the data was recorded; (ii) pointcloud datasets; (iii) range-doppler heatmaps; (iv) Annotations of the position where the subject is performing the activity (2m, 3m or 5m) and orientation with respect to the radar (left, right, front, back), activity name such as (jumping or phone typing) and activity\_class (whether macro or micro activity). These annotations are collected from the captured video frames with the help of three volunteers, who were given a remuneration of $20$ USD for each 10-minute activity session. We have summarized the collected features in \tablename~\ref{tab:features}. 

\begin{table}[]
\centering
\scriptsize
\caption{Description of features captured in the dataset.}\label{tab:features}
\scalebox{0.95}{        
\begin{tabular}{|cc|l|}
\hline
\multicolumn{2}{|c|}{\textbf{Feature}}                                        & \multicolumn{1}{c|}{\textbf{Description}}                                                                                                                                          \\ \hline
\multicolumn{2}{|c|}{datetime}                                      & \begin{tabular}[c]{@{}l@{}}The date and time when the data was recorded. This helps in time-series analysis and\\ synchronization with other data sources.\end{tabular}                           \\ \hline
\multicolumn{1}{|c|}{\multirow{7}{*}{Pointcloud}}  & rangeIdx        & \begin{tabular}[c]{@{}l@{}}Index corresponding to the range bin of the detected object. It indicates the distance of the\\ object from the radar.\end{tabular} \\ \cline{2-3} 
\multicolumn{1}{|c|}{} & dopplerIdx      & \begin{tabular}[c]{@{}l@{}}Index corresponding to the Doppler bin, which represents the relative velocity of the\\ detected object.\end{tabular}\\ \cline{2-3} 
\multicolumn{1}{|c|}{}& numDetectedObj  & \begin{tabular}[c]{@{}l@{}}The number of objects detected in a single frame. This feature is useful for understanding\\ the multi-user activity dynamics of the scene.\end{tabular}            \\ \cline{2-3} 
\multicolumn{1}{|c|}{}& range           & \begin{tabular}[c]{@{}l@{}}The actual distance measurement of the detected object from the radar in meters.\end{tabular}\\ \cline{2-3} 
\multicolumn{1}{|c|}{}                             & peakVal         & \begin{tabular}[c]{@{}l@{}}The peak value of the detected signal, indicating the strength of the returned radar signal.\end{tabular}\\ \cline{2-3} 
\multicolumn{1}{|c|}{}& x\_coord        & \begin{tabular}[c]{@{}l@{}}The x-coordinate of the detected object in the radar's coordinate system.\end{tabular}\\ \cline{2-3} 
\multicolumn{1}{|c|}{}& y\_coord        & \begin{tabular}[c]{@{}l@{}}The y-coordinate of the detected object in the radar's coordinate system.\end{tabular}\\ \hline
\multicolumn{1}{|c|}{Range-doppler Heatmap}        & doppz           & \begin{tabular}[c]{@{}l@{}}The Range-doppler Heatmap value indicating the radial velocity of the detected object,\\ helping to distinguish between stationary and moving objects.\end{tabular} \\ \hline
\multicolumn{1}{|c|}{\multirow{4}{*}{Annotations}} & Position & The position of the subject with respect to the radar. It can have values like 2m, 3m and 5m. \\ \cline{2-3}
\multicolumn{1}{|c|}{}                             & Orientation & \begin{tabular}[c]{@{}l@{}}The orientation of the subject relative to the radar's bore-sight angle: left, right, front, and back.\end{tabular}\\ \cline{2-3} 
\multicolumn{1}{|c|}{}                             & activity        & \begin{tabular}[c]{@{}l@{}}The specific activity being performed by the subject, such as walking, running, or typing,\\ used for machine learning and classification tasks.\end{tabular}       \\ \cline{2-3} 
\multicolumn{1}{|c|}{}& activity\_class & \begin{tabular}[c]{@{}l@{}}A broad categorical label of the type of activity: whether macro activity  or micro activity\end{tabular}\\ \hline
\end{tabular}}
\end{table}

\subsubsection{Dataset Access and Metadata}
Our dataset is distributed as an open-source dataset. We openly publish the dataset, along with the files for preprocessing, cleaning, and the codebase for the CNN activity classifier, along with the source code to process our data to run the existing pointcloud-based HAR activity classifier RadHAR~\cite{singh2019radhar}. Any work using our dataset should cite the main paper and the individual dataset sources listed in the documentation. The details of the codebase for data collection, pre-processing, and activity classification are provided in the GitHub repository: \url{https://github.com/arghasen10/mmDoppler}. 

\subsection{More classification report under different scenarios}

\begin{table}
	\centering
	\scriptsize
	\caption{2D-CNN architecture (M: macro, $\mu$: micro)}
	\label{tab:2d_cnn}
	\begin{tabular}{|l|c|c|c|c|c|c|} 
		\hline
		\multirow{3}{*}{\textbf{CNN Layer}} & \multicolumn{6}{c|}{Parameters}                                                           \\ 
		\cline{2-7}
		& \multicolumn{2}{c|}{Kernel} & \multicolumn{2}{c|}{Stride} & \multicolumn{2}{c|}{Channel}  \\ 
		\cline{2-7}
		& M     & $\mu$               & M     & $\mu$               & M  & $\mu$                    \\ 
		\hline
		Input Layer                         & -     & -                   & -     & -                   & 5  & 2                        \\ 
		\hline
		Conv1                               & 2 x 5 & 3 x 2               & 1 x 2 & 2 x 1               & 32 & 32                       \\ 
		\hline
		Conv2                               & 2 x 3 & 3 x 3               & 1 x 2 & 2 x 2               & 64 & 64                       \\ 
		\hline
		Conv3                               & 2 x 3 & 3 x 3               & 1 x 2 & 2 x 2               & 96 & 96                       \\ 
		\hline
		Conv4                               & 2 x 3 & -                   & 1 x 2 & -                   & 96 & -                        \\ 
		\hline
		G-avg Pool                          & -     & -                   & -     & -                   & -  & -                        \\ 
		\hline
		Softmax                             & -     & -                   & -     & -                   & 10 & 9                        \\
		\hline
	\end{tabular}
\end{table}

Existing works on HAR in mmWave domain primarily focus on pointcloud based datasets, which can only capture coarse-grained movements such as walking, running or jumping~\cite{singh2019radhar, palipana2021pantomime, alam2020lamar}, thus these existing classifiers fail to capture micro activities having subtle body movements such as phone typing, eating, etc. so in this work, we have trained the classifier with the help of denser range-doppler heatmaps captured at different doppler resolution. We leverage a 2D-CNN classifier architecture to extract the key features for different activities. The classifier architecture is shown in \tablename~\ref{tab:2d_cnn}.

Finally, we compare the proposed activity classifier with the state-of-the-art pointcloud-based classifier RadHAR~\cite{singh2019radhar}. Since RadHAR captures 3D pointclouds, however, we have collected 2D pointclouds in this work, so we have modified\footnote{\url{https://github.com/arghasen10/mmDoppler/tree/main/models}} the RadHAR classifier so that it can incorporate 2D pointclouds for the classification task.

\begin{figure*}
    \centering
    \subfloat[]
    {\includegraphics[width=0.44\textwidth]{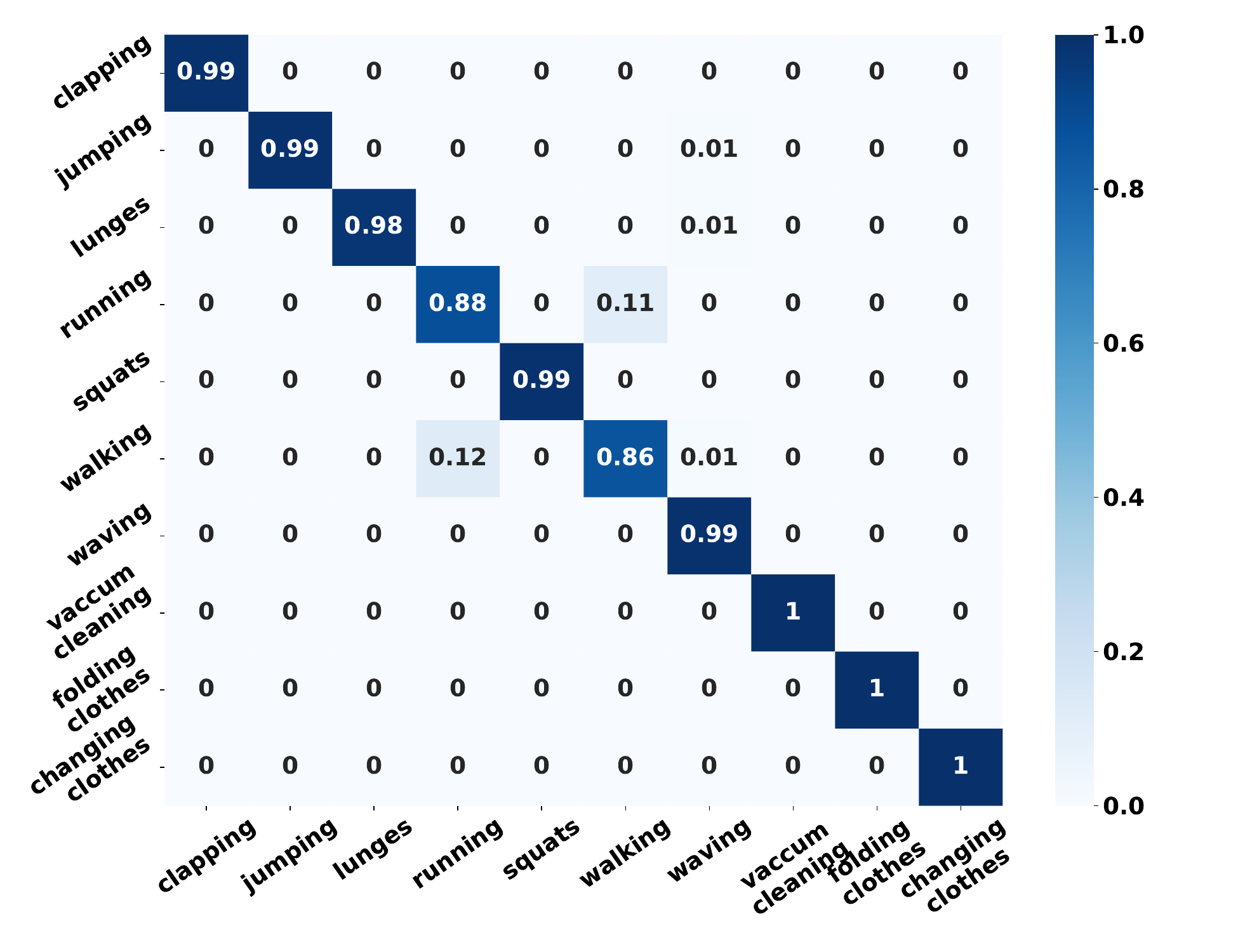}   \label{fig:mmd_macro}}
     \subfloat[]{\includegraphics[width=0.44\textwidth]{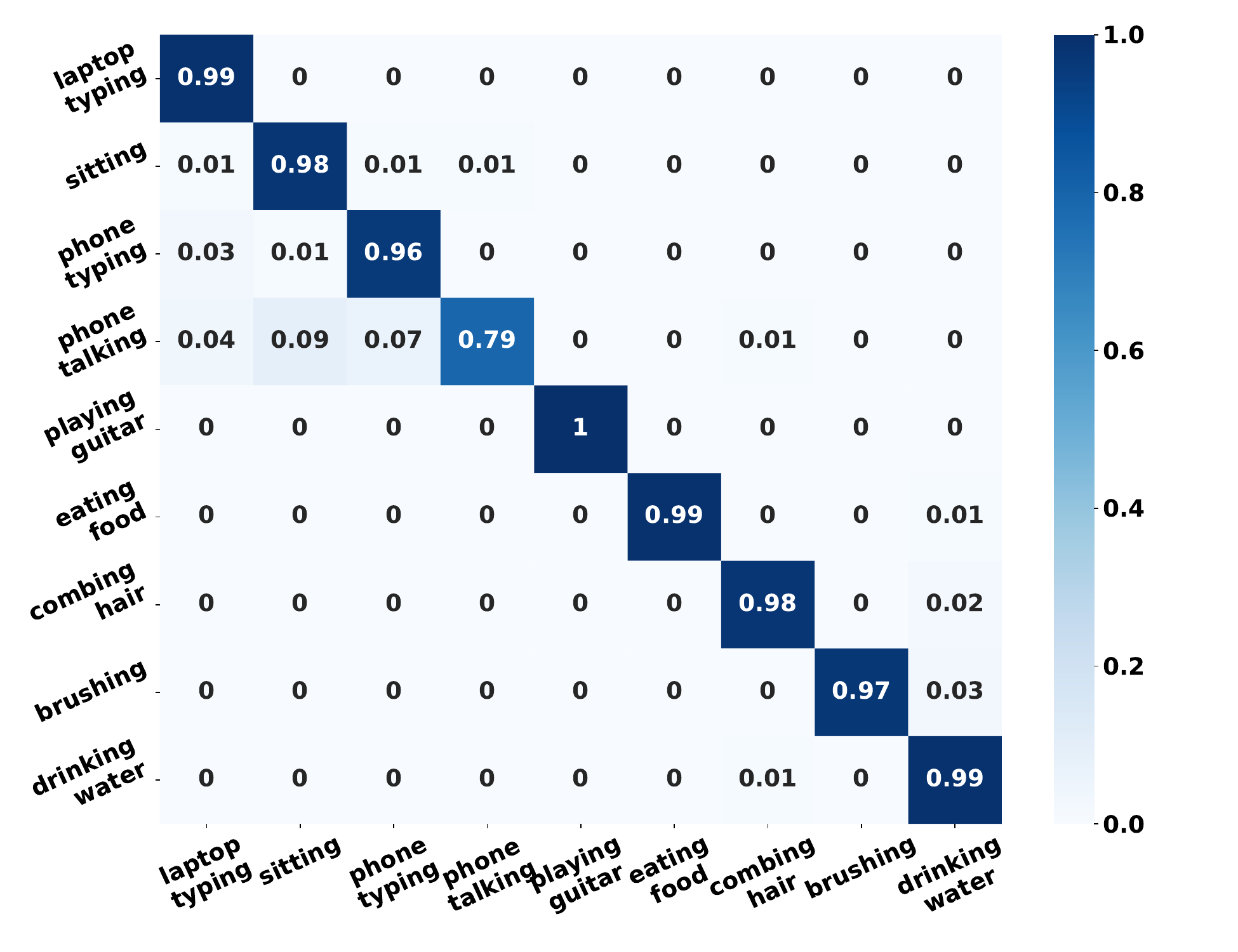}\label{fig:mmd_micro}}\\
     \subfloat[]
    {\includegraphics[width=0.44\textwidth]{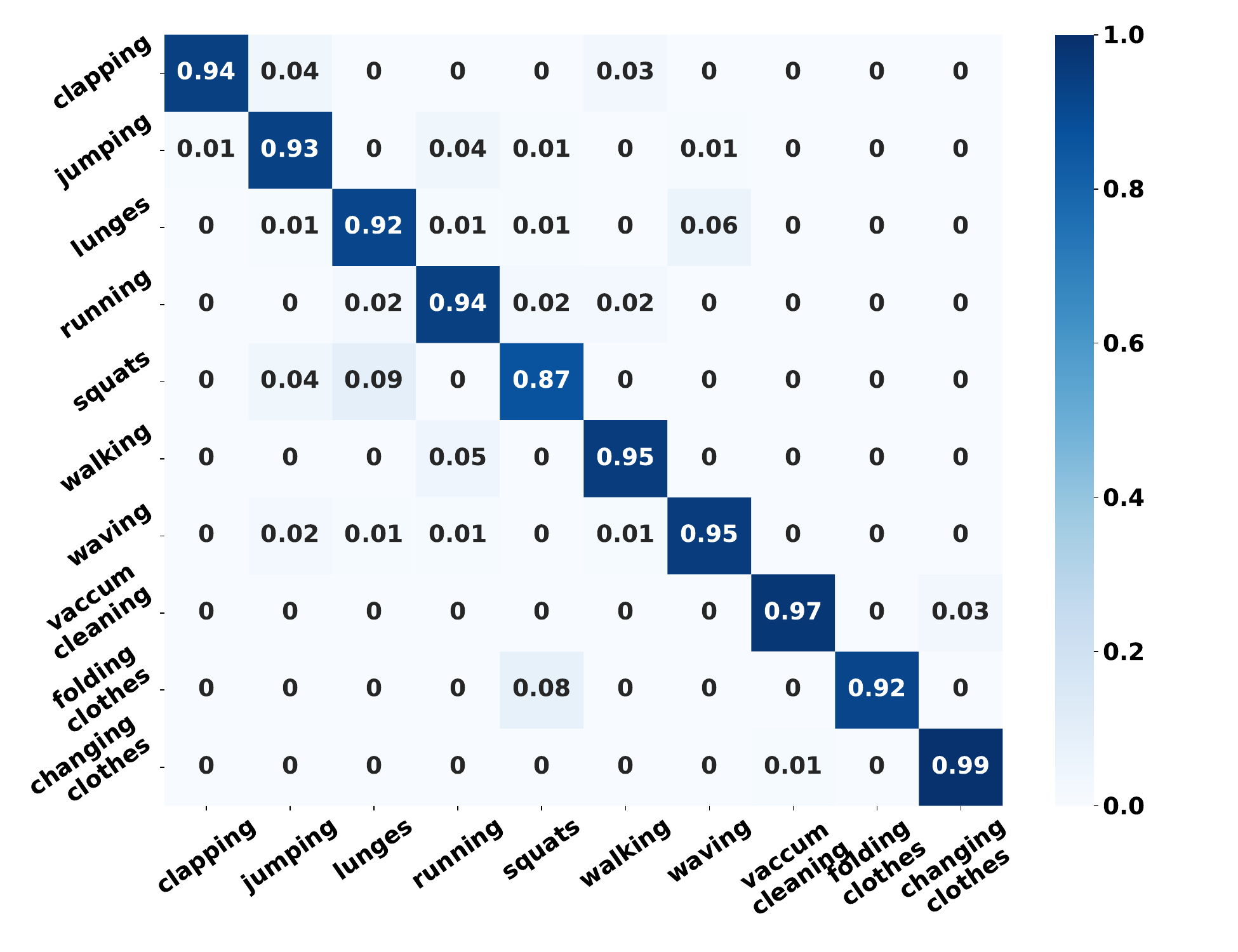}\label{fig:rad_macro}}
     \subfloat[]{\includegraphics[width=0.44\textwidth]{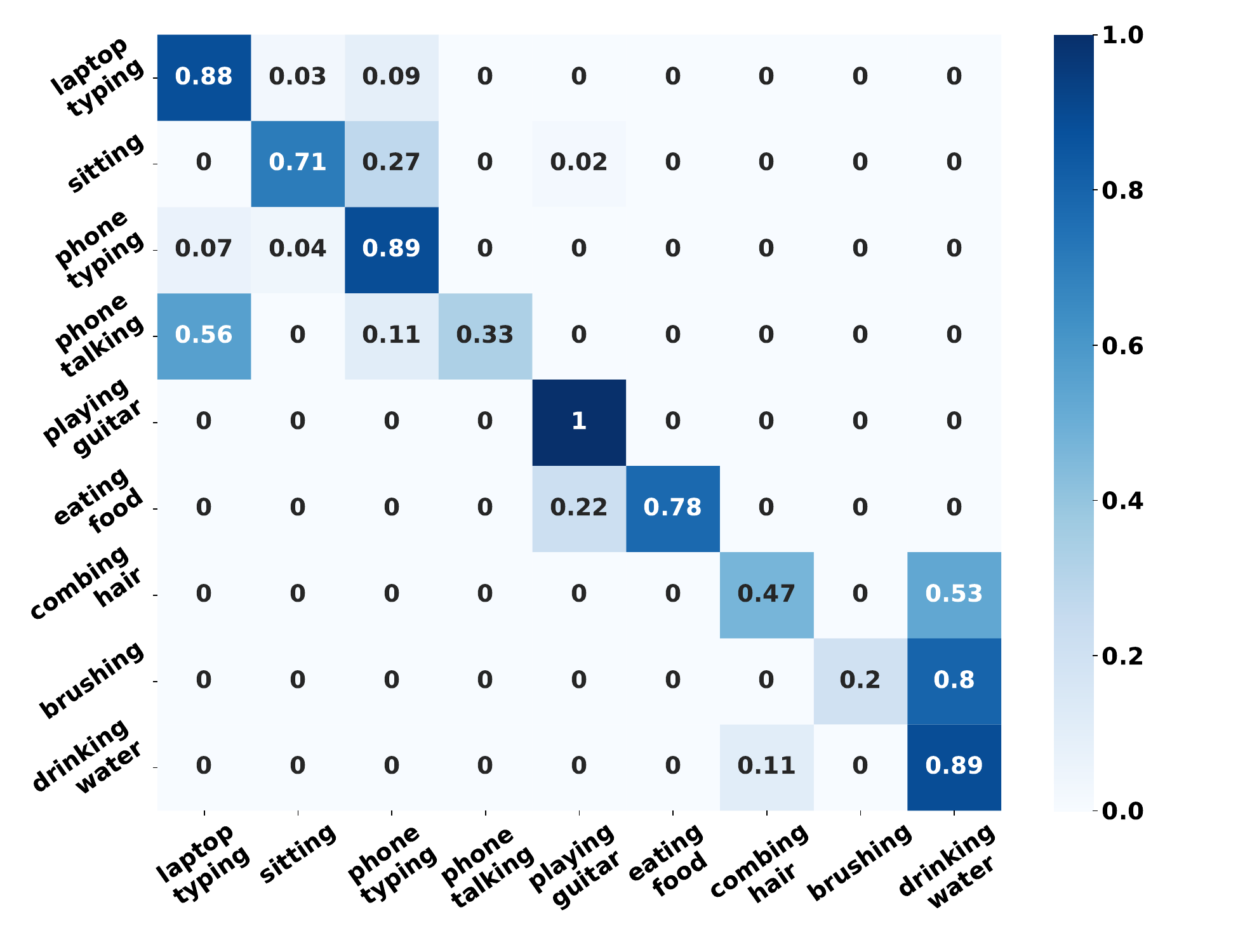}\label{fig:rad_micro}}
     \caption{Confusion matrix of the proposed classifier trained on range-doppler heatmaps for (a) macro, (b) micro activities; RadHAR trained on pointcloud data for (c) macro, (d) micro activities.}\label{fig:class}
\end{figure*}

We have illustrated the confusion matrix for activity classification using the proposed classifier, trained on range-Doppler heatmaps, and RadHAR~\cite{singh2019radhar}, trained on point cloud datasets, in \figurename~\ref{fig:class}. As depicted in \figurename~\ref{fig:mmd_macro} and \figurename~\ref{fig:mmd_micro}, the proposed 2D-CNN classifier effectively classifies activities using range-Doppler heatmaps. In contrast, the RadHAR classifier, trained on point cloud data, shows better performance in classifying macro activities involving significant movements (see \figurename~\ref{fig:rad_macro}). However, it struggles with micro activities that involve subtle body movements (see \figurename~\ref{fig:rad_micro}). This is primarily because existing methods and datasets are influenced by pointcloud datasets which although can work on macro activities but fail miserably on micro activities. Thus, our dataset containing range-Doppler heatmaps, with their denser representation, can capture features more accurately than the sparser point cloud representations.

\end{acronym}
\end{document}